\documentclass[a4paper,UKenglish,cleveref, autoref, thm-restate]{lipics-v2021}

\title{Analytical Differential Calculus with Integration} 

\titlerunning{Analytical Differential Calculus with Integration} 

\author{Han Xu}{Department of Computer Science and Technology\\Peking University, China}{1800012917@pku.edu.cn}{}{}
\author{Zhenjiang Hu}{Key Laboratory of High Confidence Software Technologies (MoE)\\Department of Computer Science and Technology\\Peking University, China}{huzj@pku.edu.cn}{}{}

\authorrunning{H. Xu and Z. Hu} 

\Copyright{Han Xu and Zhenjiang Hu} 




\category{} 

\relatedversion{} 

\nolinenumbers 

\EventEditors{}
\EventNoEds{2}
\EventLongTitle{448th International Colloquium on Automata, Languages, and Programming (ICALP 2021)}
\EventShortTitle{ICALP 2021}
\EventAcronym{ICALP}
\EventYear{2021}
\EventDate{July 12--16, 2021}
\EventLocation{Glasgow, Scotland}
\EventLogo{}
\SeriesVolume{42}
\ArticleNo{23}
\usepackage{booktabs}   
\usepackage{subcaption} 
\usepackage{bussproofs}
\usepackage{bcprules}
\usepackage{float}
\usepackage{amsmath}
\usepackage{amssymb}
\usepackage{color}
\usepackage{setspace}

\newcommand\m[1]{\mbox{\it #1}}

\newcommand\mynote[1]{\mbox{\small #1}}

\def\nf{\m{nf}}
\def\nb{\m{nb}}
\def\itm{\m{it}}
\def\iB{\m{iB}}

\newcommand{\myder}[3]{\frac{\partial #1}{\partial #2}|_{#3}}
\newcommand{\myint}[3]{\int_{#1}^{#2} #3\ dx}
\newcommand{\mycase}[5]{case~#1~of~inl~#2\Rightarrow #3~|~inr~{#4}\Rightarrow #5 }
\newcommand{\tder}[2]{\frac{\partial #1}{\partial #2}}
\newcommand{\red}[2]{#1\,\rightarrow\,#2}
\def\myend{\flushright{$\blacksquare$}}
\def\basetype{\textsf{B}}
\newcommand\ignore[1]{}
\newcommand\tjudge[3]{#1\,\vdash\,#2:\,#3}
\newcommand\reason[1]{\mbox{\footnotesize \quad\{~#1~\}}}

\def\reductionc{
\begin{prooftree}
\AxiomC{$t_0:\basetype $}
\UnaryInfC{$\myder{ (t_1, t_2,...,t_{n})}{x}{t_0}\ \rightarrow \ (\myder{t_1}{x}{t_0},\myder{t_2}{x}{t_0},...,\myder{t_{n}}{x}{t_0})
$}
\end{prooftree}
}
\def\reductiond{\begin{prooftree}
\AxiomC{$t_0:\basetype $}
\UnaryInfC{$ \myder{(\lambda y:T. t)}{x}{t_0}\ \rightarrow \ \lambda y:T . \myder{t}{x}{t_0}
$}
\end{prooftree}
}
\def\reductione{
\begin{prooftree}
\AxiomC{$t_1,t_2:\basetype $}
\UnaryInfC{$\int_{t_1}^{t_2}(t_{11},t_{12},...t_{1n})dx~ \rightarrow (\int_{t_1}^{t_2}t_{11}dx,
\int_{t_1}^{t_2}t_{12}dx,...,\int_{t_1}^{t_2}t_{1n}dx)
$}
\end{prooftree}
}

\def\reductionf{
\begin{prooftree}
\AxiomC{$t_1,t_2:\basetype $}
\UnaryInfC{$\int_{t_1}^{t_2}\lambda y:T_2 .tdx~ \rightarrow \lambda y:T_2.\int_{t_1}^{t_2}t dx $}
\end{prooftree}
}

\def\reductiong{\begin{prooftree}
\AxiomC{$\forall i,(t_1, t_2...,t_{i-1},x_i,t_{i+1}...,t_{n})is~written~as~ t_{i*}$}
\UnaryInfC{ $\myder{t}{x}{(t_1, t_2,...,t_{n})} \rightarrow (\myder{t[t_{1*}/x]}{x_1}{t_1},\myder{t[t_{2*}/x]}{x_2}{t_2},...,\myder{t[t_{n*}/x]}{x_{n}}{t_{n}})$}
\end{prooftree}
}

\def\reductionh{$\int_{(t_{11},...,t_{1n}) }^{(t_{21},...,t_{2n})}tdx~ \rightarrow \int_{t_{11}}^{t_{21}}  \pi_1(t[(x_1,t_{12},...t_{1n})/x])dx_1 \oplus
\cdots \oplus \int_{t_{1n}}^{t_{2n}} \pi_n(t[(t_{21},t_{22},...,x_{n})/x])dx_{n} $
}

\def\reductionj{$(\lambda x: T.t_1) \oplus (\lambda y: T.t_2)  \rightarrow \lambda x: T.t_1 \oplus  (t_2[x/y]) $
}

\def\reductionk{$(t_{11},t_{12},...t_{1n}) \ominus (t_{21},t_{22},...t_{2n})  \rightarrow (t_{11}\ominus t_{21},t_{12}\ominus t_{22},...t_{1n}\ominus t_{2n})$
}

\def\reductionl{$(\lambda x: T.t_1) \ominus (\lambda y: T.t_2)  \rightarrow \lambda x: T.t_1 \ominus  (t_2[x/y]) $
}

\def\reductionm{\begin{prooftree}
\AxiomC{$ t_2 :\basetype $}
\UnaryInfC{$ (\lambda x:T .t) * t_2\rightarrow \lambda x:T .(t * t_2)$}
\end{prooftree}

}

\def\reductionn{
\begin{prooftree}
\AxiomC{$ t_0 :\basetype $}
\UnaryInfC{$ (t_1,t_2,...t_{n}) * t_0\rightarrow (t_1*t_0,t_2*t_0,...t_{n}*t_0)$}
\end{prooftree}

}

\def\reductiono{
\begin{prooftree}
\AxiomC{$ t_1:(t_{11},t_{12},...t_{1n}),t_2 :(t_{21},t_{22},...t_{2n}) $}
\UnaryInfC{$t_1 * t_2\rightarrow (t_{11}*t_{21})\oplus (t_{12}*t_{22})\oplus ... \oplus (t_{1n}*t_{2n}) $}
\end{prooftree}
}

\def\reductionr{
\begin{prooftree}
\AxiomC{$ t_0 :\basetype $}
\UnaryInfC{$\myder{inl/inr~t}{x}{t_0}\rightarrow inl/inr~\myder{t}{x}{t_0}$}
\end{prooftree}
}

\def\reductions{
\begin{prooftree}
\AxiomC{$ t_1,t_2 :\basetype $}
\UnaryInfC{$\int_{t_1}^{t_2}inl/inr~t~dx\rightarrow inl/inr~\int_{t_1}^{t_2}t~dx$}
\end{prooftree}
}

\def\reductiont{\red{case~(inl~t)~of~inl~{x_1}\Rightarrow t_1~|~inr~{x_2}\Rightarrow t_2 }{t_1[t/x_1]}
}

\def\reductionu{\red{case~(inr~t)~of~inl~{x_1}\Rightarrow t_1~|~inr~{x_2}\Rightarrow t_2 }{t_2[t/x_2]}
}

\def\reductionv{\begin{prooftree}
\AxiomC{$ t_0 :\basetype $}
\UnaryInfC{$ (inl/inr~t) * t_0\rightarrow inl/inr~(t * t_0)$}
\end{prooftree}

}

\def\modify#1#2#3{{\small\underline{\sf{#1}}:} {\color{red}{\small #2}}
{{\color{red}\mbox{$\Rightarrow$}}} {\color{blue}{#3}}}

\newcommand{\xmodify}[2]{#2}

\begin{document}
\maketitle




\begin{abstract}
Differential \xmodify{lambda calculus}{lambda-calculus} was first introduced by Thomas Ehrhard and Laurent Regnier in 2003. Despite more than 15 years of history, little work has been done on  a differential calculus with integration. In this paper, we shall propose a differential calculus with integration from a programming point of view. We show its good correspondence with mathematics, which is manifested by how we construct these reduction rules and how we preserve important mathematical theorems in our calculus. Moreover, we highlight applications of the calculus in incremental computation, automatic differentiation, and computation approximation.
\end{abstract}


\maketitle

\section{Introduction}







Differential calculus has more than 15 years of history in computer science since the pioneer work by Thomas Ehrhard and Laurent Regnier \cite{EhRe03}. It is, however, not well-studied from the perspective of programming languages; we would expect the profound connection of differential calculus with important fields such as incremental computation, automatic differentiation and self-adjusting computation just like how mathematical analysis connects with mathematics.\xmodify{Review 3 want us to clarify what kind of connection is here, but I feel it is unrelated to the mainline.}{} We want to understand what is the semantics of the derivative of a program and how we can use these derivatives to write a program. \xmodify{Technically}{That is}, we wish to have a clear description of derivatives and introduce integration to compute from operational derivatives to the program.

The two main lines of the \xmodify{none}{related} work are the differential lambda-calculus \cite{EhRe03,Ehrh18} and the change theory \xmodify{\cite{CGRO14,Alva19},Added Reference : The Difference Lambda-Calculus: A Language for Difference Categories, as suggested by Review 2}{\cite{CGRO14,Alva19,Alvar20}}. On one hand, the differential lambda-calculus uses linear substitution to represent the derivative of a term. For example, given a term $x*x$ (i.e., $x^2$), with the differential lambda-calculus, we may use the term $\frac{\partial x*x}{\partial x}\cdot 1$ to denote its derivative at $1$. As there are two alternatives to substitute $1$ for $x$ in the term $x*x$, it gives $(1*x)+(x*1)$ (i.e., $2 x$) as the derivative (where $+$ denotes "choice"). \xmodify{Review 1 suggests Church addition $\lambda f~g ~x.f~ (g~ x)$ should be used here, but I feel it is unnecessary}{}

Despite that the differential lambda-calculus provides a concise way to analyze the alternatives of linear substitution on a lambda term, there is a gap between analysis on terms and computation on terms. For instance, let $+'$ \xmodify{denotes}{denote} our usual addition operator, \xmodify{none}{and $+$ denote the choice of linear substitution.} \xmodify{We will have}{Then we have that} $\frac{\partial x +' x}{\partial x}\cdot 1$ = $(1 +' x)+(x +' 1)$, which is far away from the expected $1 +' 1$. Moreover, it offers no method to integrate over a derivative, say $\frac{\partial t}{\partial x}\cdot y$.

On the other hand, the change theory gives a systematic way to define and propagate (transfer) changes. The main idea is to define the change of function $f$ as $\m{Derive}~f$, satisfying
\[
f (x \oplus \Delta x) = f(x) \oplus (\m{Derive}~f)~x~\Delta x.
\]
where $\oplus$ denotes an updating operation. It reads that the change over the input $x$ by $\Delta x$ results in the change over the result of $f(x)$ by $(\m{Derive}~f)~x~\Delta x$.
 While \xmodify{the change theory}{change theory} provides a general way to describe changes, the changes it described are differences (deltas) instead of derivatives. It is worth noting that derivative is not the same as delta. For example, by \xmodify{the change theory}{change theory}, we can deduce that $f(x)$ will be \xmodify{in}{of} the form of $x*x+C$ if we know $(\m{Derive}~f)~x~\Delta x = 2 * x * \Delta x + \Delta x * \Delta x$, but we cannot deduce this form if we just know that its derivative is $2*x$, because \xmodify{the change theory}{change theory} has no concept of integration or \xmodify{limitation}{limits}.

Although a bunch of work has been done on derivatives \xmodify{\cite{EhRe03,Ehrh18,CGRO14,Alva19,PaKo82,LaZh07,SiPe08,Elli18} (Added Reference : A Simple Differentiable Programming Language as suggested by Review 2)}{\cite{EhRe03,Ehrh18,CGRO14,Alva19,PaKo82,LaZh07,SiPe08,Elli18,Abadi20}}, there is unfortunately, as far as we are aware, little work on integration. It may be natural to ask what a derivative really means if we cannot integrate it. If there is only a mapping from a term to its derivative without its corresponding integration, how can we operate on derivatives with a clear understanding of what we actually have done?
\xmodify{Review 1 argues that some differentiation have their meaning even without integration, Review 2 casr doubt on why it is natural to ask the integration. I think it makes some points, but I am not sure how to modify it.}{}

In this paper, we aim at a new differential framework, having dual mapping between derivatives and integrations. With this framework, we can manifest the power of this dual mapping by proving, among others, three important theorems, namely the Newton-Leibniz formula, the Chain Rule and the Taylor's theorem.

Our key idea can be illustrated by a simple example. Suppose we have a function $f$ mapping from an $n$-dimensional space to an $m$-dimensional space. Then, let $x$ be $(x_1,x_2,...,x_n)^T$, and $f(x)$ be $(f_1(x),f_2(x),...,f_m(x))^T$.
%
Mathematically, we can use a \xmodify{none}{Jacobian} matrix $A$ to represent its derivative, which satisfies the equation
\[
f (x + \Delta x) - f(x) = A \Delta x + o(\Delta x)
,~\xmodify{where}{\text{where}}~A~=
\begin{pmatrix}
  \frac{\partial f_1}{\partial x_1} & \frac{\partial f_1}{\partial x_2} & \cdots & \frac{\partial f_1}{\partial x_n}\\
  \frac{\partial f_2}{\partial x_1} & \frac{\partial f_2}{\partial x_2} & \cdots & \frac{\partial f_2}{\partial x_n}\\
  \cdots &\cdots &\cdots &\cdots \\
  \frac{\partial f_m}{\partial x_1} & \frac{\partial f_m}{\partial x_2} & \cdots & \frac{\partial f_m}{\partial x_n}
\end{pmatrix}
\]
However, computer programs usually describe computation over data of some structure, rather than just scalar data or matrix. In this paper, we extend the idea and propose a new calculus that enables us to perform differentiation and integration on data structures. Our main contributions are summarized as follows.
\begin{itemize}
\item \xmodify{none}{To our knowledge}, we have made the first attempt of designing a calculus that provides both derivative and integral. It is an extension of the \xmodify{lambda calculus}{lambda-calculus} with five new operators including derivatives and integrations. We give clear semantics and typing rules, and prove that it is sound and strongly normalizing. (Section \ref{sec:calculus})

\item We prove three important theorems and highlight their practical application for incremental computation, automatic differentiation, and computation approximation.

\begin{itemize}
\item We prove the Newton-Leibniz formula: $\int_{t_1}^{t_2} \myder{t}{y}{x} dx = t[t_2/y]\ominus t[t_1/y]$, \xmodify{Suggested by Review 2}{ which is also known as Second Fundamental Theorem of Calculus.} It shows the duality between derivatives and integrations, and can be used for incremental computation. (Section \ref{sec:NL})

\item We prove the Chain Rule:
$\myder{f(g~x)}{x}{t_1}*t=\myder{f~y}{y}{g~t_1}*(\myder{g~z}{z}{t_1}*t)$. It says $\forall x,\forall x_0, (f(g(x)))'*x_0=f'(g(x))*g'(x)*x_0$, and can be used for incremental computation and automatic differentiation. (Section \ref{sec:chain})

\item We prove the Taylor's Theorem: $f~ t = \sum\limits_{k=0}^{\infty}\frac{1}{k!}(f^{(k)}~t_0)*(t\ominus t_0)^k $. Different from that one of the differential lambda-calculus \cite{EhRe03}, this Taylor's theorem manifests results of computation instead of analysis on occurrence of terms. It can be used for approximation of a function computation. (Section \ref{sec:taylor})
\end{itemize}

\end{itemize}


\section{Calculus}
\label{sec:calculus}

In this section, we shall give a clear definition of our calculus with both derivatives and integration.  We explain important insights in our design, and prove some useful properties and theorems that will be used later.

\subsection{Syntax}

\begin{figure}[t]
\[
\begin{array}{lrcll}
\mbox{Terms} & t &::=
 & c & \mynote{constants of interpretable type}\\
 && | & x           & \mynote{variable}\\
 && | & \lambda x:T.\,t & \mynote{lambda abstraction}\\
 && | & t\ t  & \mynote{function application}\\
 && | & (t_1,t_2, \dots,t_{n})~|~\pi_j~t & \mynote{$n$-tuple and projection}\\
 && | & t \oplus t & \mynote{addition}\\
 && | & t \ominus t & \mynote{subtraction}\\
 && | & t * t       & \mynote{multiplication}\\ \vspace{1ex}
 && | & \myder{t}{x}{t} & \mynote{derivative\xmodify{fix the line space}{}}\\
 && | & \myint{t}{t}{t} & \mynote{integration}\\
 && | & inl~t ~|~ inr~t & \mynote{left/right injection}\\
 && | & \mycase{t}{x_1}{t}{x_2}{t} & \mynote{case analysis}\\
 && | & fix~t & \mynote{fix point}\\

 \\
 \mbox{Types} & T &::=& \basetype & \mynote{base type}\\
 && | & (T_1,T_2,\ldots,T_{n}) & \mynote{product type}\\
 && | & T \rightarrow T & \mynote{function type}\\
 && | & T + T & \mynote{sum type}\\

 \\
 \mbox{Contexts} & \Gamma &::=& \emptyset & \mynote{empty context}\\
  && | & \Gamma,\, x:T & \mynote{variable binding}
\end{array}
\]
\caption{Calculus Syntax}
\label{fig:calculus}
\end{figure}

Our calculus, as defined in Figure \ref{fig:calculus}, is an extension of the simply-typed lambda calculus \cite{Benj02}. Besides the usual constant, variable, lambda abstraction, function application, and tuple, it introduces five new operations: addition $\oplus$, subtraction $\ominus$, multiplication $*$, derivative $\myder{t}{x}{t}$ and integration $\myint{t}{t} {t}$. The three binary operations, namely $\oplus$, $\ominus$, and $*$, are generalizations of those from our mathematics. Intuitively, $x \oplus \Delta$ is for updating $x$ with change $\Delta$, $\ominus$ for canceling updates, and \xmodify{$\otimes$}{*} for distributing updates. We build up terms from terms of base types (such as $\mathbb{R}$, $\mathbb{C}$), and on each base type we require these operations satisfy the following properties:
\begin{itemize}
\item The addition and multiplication are associative and commutative, i.e., $(a \oplus b) \oplus c = a \oplus (b \oplus c)$, $a \oplus b = b \oplus a$, $(a * b) * c = a * (b * c)$, $a * b = b * a$.
\item The addition and the subtraction are cancellable, i.e., $(a \oplus b) \ominus b = a$ and $(a \ominus b) \oplus b = a$.
\item The multiplication is distributive over addition, i.e., $ a * (b \oplus c) = a * b \oplus a * c$.
\end{itemize}

\begin{example}[Basic Operations on Real Numbers]
\rm
For real numbers $r_1, r_2 \in \mathbb{R}$, we have the following definitions.
\[
\begin{array}{llll}
r_1 \oplus r_2 &=& r_1 + r_2\\
r_1 \ominus r_2 &=& r_1 - r_2\\
r_1 * r_2 &=& r_1\ r_2
\end{array}
\]
\end{example}

We use $\myder{t_1}{x}{t_2}$ to denote derivative of $t_1$ over $x$ at point $t_2$, and $\myint{t_1}{t_2} {t}$ to denote integration of $t$ over $x$ from $t_1$ to $t_2$.

\subsection{Typing}

\begin{figure}[tb]

\begin{center}

\begin{minipage}{0.45\hsize}
\infrule[TCon]
{c:T \in \Gamma}
{\tjudge{\Gamma}{c}{T}}
\end{minipage}
\qquad
\begin{minipage}{0.45\hsize}
\infrule[TVar]
{x:T \in \Gamma}
{\tjudge{\Gamma}{x}{T}}
\end{minipage}
\\

~\\
\begin{minipage}{0.45\hsize}
\infrule[TInl]
{\tjudge{\Gamma}{t}{T_1}}
{\tjudge{\Gamma}{inl~t}{T_1+T_2}}
\end{minipage}
\quad
\begin{minipage}{0.45\hsize}
\infrule[TInr]
{\tjudge{\Gamma}{t}{T_2}}
{\tjudge{\Gamma}{inr~t}{T_1+T_2}}
\end{minipage}
\\
~\\
\begin{minipage}{0.45\hsize}
\infrule[TAdd]
{\tjudge{\Gamma}{t_1}{T^*} \quad \tjudge{\Gamma}{t_2}{T^*}}
{\tjudge{\Gamma}{t_1 \oplus t_2}{T^*}}
\end{minipage}
\quad
\begin{minipage}{0.45\hsize}
\infrule[TSub]
{\tjudge{\Gamma}{t_1}{T^*} \quad \tjudge{\Gamma}{t_2}{T^*}}
{\tjudge{\Gamma}{t_1 \ominus t_2}{T^*}}
\end{minipage}
\\
~\\
\begin{minipage}{0.45\hsize}
\infrule[TAbs]
{\tjudge{\Gamma, x:T_1}{t}{T_2}}
{\tjudge{\Gamma}{\lambda x:T_1.\,t}{T_1\rightarrow T_2}}
\end{minipage}
\qquad
\begin{minipage}{0.45\hsize}
\infrule[TApp]
{\tjudge{\Gamma}{t_1}{T_1 \to T_2} \quad \tjudge{\Gamma}{t_2}{T_1} }
{\tjudge{\Gamma}{t_1~t_2}{T_2}}
\end{minipage}
\\
~\\
\begin{minipage}{0.45\hsize}
\infrule[TFix]
{\tjudge{\Gamma}{t}{T\rightarrow T}}
{\tjudge{\Gamma}{fix~t}{T}}
\end{minipage}
\quad
\begin{minipage}{0.45\hsize}
\infrule[TDer]
{\tjudge{\Gamma}{t_1}{T_1} \quad \tjudge{\Gamma,x:T_1}{t_2}{T_2}}
{\tjudge{\Gamma}{\myder{t_2}{x}{t_1}}{\tder{T_2}{T_1}}}
\end{minipage}
\\
~\\
\begin{minipage}{0.45\hsize}
\infrule[TPair]
{\tjudge{\forall j\in[1,n], \Gamma}{t_j}{T_j} }
{\tjudge{\Gamma}{(t_1,t_2,...,t_{n})}{(T_1,T_2,...,T_{n})}}
\end{minipage}
\quad
\begin{minipage}{0.45\hsize}
\infrule[TProj]
{\tjudge{\forall j\in[1,n],\Gamma}{t}{(T_1,T_2,...,T_{n})}}
{\tjudge{\Gamma}{\pi_j~ t}{T_j}}
\end{minipage}
\\

~\\
\begin{minipage}{0.85\hsize}
\infrule[TMul]
{\tjudge{\Gamma}{t_1}{\tder{T^*}{T}} \quad \tjudge{\Gamma}{t_2}{T}}
{\tjudge{\Gamma}{t_1 * t_2}{T^*}}
\end{minipage}
\\
\begin{minipage}{0.85\hsize}
\infrule[TInt]
{\tjudge{\Gamma}{t_1}{T} \quad \tjudge{\Gamma}{t_2}{T} \quad \tjudge{\Gamma,x:T}{t}{\tder{T^*}{T}}}
{\tjudge{\Gamma}{\myint{t_1}{t_2}{t}}{T^*}}
\end{minipage}
\\
~\\
\begin{minipage}{0.85\hsize}
\infrule[TCase]
{\tjudge{\Gamma,x_1:T_1}{t_1}{T} \quad \tjudge{\Gamma,x_2:T_2}{t_2}{T}\quad \tjudge{\Gamma}{t}{T_1+T_2}}
{\tjudge{\Gamma}{\mycase{t}{x_1}{t_1}{x_2}{t_2}}{T}}
\end{minipage}

\end{center}
\caption{Typing Rules}
\label{fig:typing}
\end{figure}

\xmodify{As stated by Review 3, TMul2 and TInt2 are redundant because they are special case of TMul1 and TInt1, so I delete them}{}

As defined in Figure \ref{fig:calculus}, we have base types (denoted by $\basetype$), tuple types, function types, and sum type. To make our later typing rules easy to understand, we introduce the following type notations.
\[
\begin{array}{lrcll}
\mbox{Type} & T^* &::=
 & B & \mynote{base type}\\
 && | & (T^*,T^*,...,T^*)   & \mynote{product type}\\
 && | & T\rightarrow T^* & \mynote{arrow type}\\
\end{array}
\]

$T^*$ means the types that are addable (i.e., updatable through $\oplus$). We view the addition between functions, tuples and base type terms as valid, which will be showed by our reduction rules later. But here, we forbid the addition and subtraction between sum types because we view updates such as $inl~0\oplus inr~1$ as invalid. If we want to update the change to a term of sum types anyway, we may do case analysis such as  $\mycase{t}{x_1}{inl~(x_1\oplus...)}{x_2}{(x_2\oplus...)}$.

Next, we introduce two notations for derivatives on types:
\[
\frac{\partial T}{\partial \basetype} = T,
\]
\[
\frac{\partial T}{\partial (T_1,T_2,...,T_{n})} =(\frac{\partial T}{\partial T_1},\frac{\partial T}{\partial T_2},...,\frac{\partial T}{\partial T_{n}}).
\]
The first notation says that with the assumption that differences (subtraction) of values of base types are of base types, the derivative over base types has no effect on the result type.
And, the second notation resembles partial differentiation. Note that we do not consider derivatives on functions because even for functions on real numbers, there is no good mathematical definition for them yet. Therefore, we do not have a type notation for $\frac{\partial T}{\partial (T_1\rightarrow T_2)}$. Besides, because we forbid the addition and subtraction between the sum types, we will iew the differentiation of the sum types as invalid, so we do not have notations for $\frac{\partial T}{\partial (T_1+ T_2)} $ either.

Figure \ref{fig:typing} shows the typing rules for the calculus. The typing rules for constant, variable, lambda abstraction, function application, tuple, and projection are nothing special. The typing rules for addition and subtraction are natural, but the rest three kinds of rules are more interesting.
\xmodify{Rule {\sc TMul1} and {\sc TMul2} show}{Rule {\sc TMul}}  the typing rule for $t_1 * t_2$. If $t_1$ is a derivative of $T_1$ over $T_2$, and $t_2$ is of type $T_2$, then multiplication will produce a term of type $T_1$. This may be informally understood from our familiar equation $\frac{\triangle{Y}}{\triangle{X}} * \triangle{X} = \triangle{Y}$.
Rule {\sc TDer} shows introduction of the derivative type through a derivative operation, while
\xmodify{Rule {\sc TInt1} and {\sc TInt2} show}{Rule {\sc TInt}}  cancellation of the derivative type through an integration operation.

\subsection{Semantics}

We will give a two-stage semantics for the calculus. At the first stage, we assume that all the constants (values and functions) over the base types are {\em interpretable} in the sense there is a default well-defined interpreter to evaluate them. At the second stage, the important part of this paper, we define a set of reduction rules and use the full reduction strategy to compute their normal form, which enjoys good properties of soundness, confluence, and strong normalization.

More specifically, after the full reduction of a term in our calculus, every subterm (now in a normal form of interpretable types) outside the lambda function body will be interpretable on base types, which will be proved in the appendix. In other words, our calculus helps to reduce a term to a normal form which is interpretable on base types, and leave the remaining evaluations to interpretation on base types. We will not give reduction rules to the operations on base types because we do not want to touch on implementations of primitive functions on base types.

For simplicity, in this paper we will assume that the important properties such as the Newton-Leibniz formula, the Chain Rule, and the Taylor's theorem, are satisfied by all the primitive functions and their closures through addition, subtraction, multiplication, derivative and integration.
This assumption may seem too strong, since not all primitive functions on base types meet this assumption. However, it would make sense to start with the primitive functions meeting these requirements to build our system, and extend it later with other primitive functions.

\subsection{Interpretable Types and Terms}

Here, a term is interpretable means it can be directly interpreted by \xmodify{base type interpreter}{a base type interpreter}. We use $\basetype$ to denote the base type, over which its constants are interpretable.    To make this clear, we define interpretable types as follows.

\begin{definition}[Interpretable Type]\rm
  Let $\basetype$ be base types. A type $\iB$ is interpretable if it is generated by the following grammar:
\[
\begin{array}{llll}
  \iB &::=& \basetype & \mynote{base type}\\
     & | & \iB \rightarrow \iB & \mynote{function type}
\end{array}
\]
\end{definition}

Constants of interpretable types can be both values or primitive functions of base types. For example, we can use $\sin(x)$, $\cos(x)$, $\m{square}(x)$ as primitive functions in our calculus.

Next, we consider terms that are constructed from constants and variables of interpretable types. These terms are interpretable by a default evaluator under an environment mapping variables to constants.
Formally, we define the following interpretable terms.

\begin{definition}[Interpretable Terms]\rm A term is an interpretable if it belongs to $\itm$.
\[
\begin{array}{lrcll}
  & \itm &::=
 & c & \mynote{constants of $\iB$}\\
 && | & x           & \mynote{variable of $\iB$ }\\
 && | & \lambda x:\iB.\,\itm & \mynote{lambda abstraction}\\
 && | & \itm\ \itm  & \mynote{function application}\\
 && | & \itm \oplus \itm & \mynote{addition}\\
 && | & \itm \ominus \itm & \mynote{subtraction}\\
 && | & \itm * \itm       & \mynote{multiplication}\\\vspace{1ex}

 && | & \myder{\itm}{x}{\itm} & \mynote{derivative}\xmodify{fix the space}{}\\
 && | & \myint{\itm}{\itm}{\itm} & \mynote{integration}\\
\end{array}
\]
\end{definition}

\subsection{Reduction Rules}

\begin{figure}[tb]
\centering

\begin{minipage}{0.90\hsize}
\infrule[EAppDer1]
{t_0:\basetype }
{\red{\myder{ (t_1, t_2,...,t_{n})}{x}{t_0}}
     { (\myder{t_1}{x}{t_0},\myder{t_2}{x}{t_0},...,\myder{t_{n}}{x}{t_0})}}
\end{minipage}
\\
~\\
\begin{minipage}{0.90\hsize}
\infrule[EAppDer2]
{t_0:\basetype}
{\red{\myder{inl/inr~t}{x}{t_0}}{inl/inr~\myder{t}{x}{t_0}}}
\end{minipage}
\\
~\\
\begin{minipage}{0.90\hsize}
\infrule[EAppDer3]
{t_0:\basetype }
{\red{\myder{(\lambda y:T. t)}{x}{t_0}}
     {\lambda y:T . \myder{t}{x}{t_0}}}
\end{minipage}
\\
~\\
\begin{minipage}{0.90\hsize}
\infrule[EAppDer4]
{\forall i\in[1,n],\, t_{i*} = (t_1, t_2...,t_{i-1},x_i,t_{i+1}...,t_{n})}
{\red{\myder{t}{x}{(t_1, t_2,...,t_{n})}}
     {(\myder{t[t_{1*}/x]}{x_1}{t_1},\myder{t[t_{2*}/x]}{x_2}{t_2},...,\myder{t[t_{n*}/x]}{x_{n}}{t_{n}})}}
\end{minipage}
\\
~\\
\begin{minipage}{0.90\hsize}
\infrule[EAppInt1]
{t_1,t_2:\basetype }
{\red{\int_{t_1}^{t_2}(t_{11},t_{12},...t_{1n})dx}
     {  (\int_{t_1}^{t_2}t_{11}dx,
\int_{t_1}^{t_2}t_{12}dx,...,\int_{t_1}^{t_2}t_{1n}dx)}}
\end{minipage}
\\
~\\
\begin{minipage}{0.90\hsize}
\infrule[EAppInt2]
{t_1,t_2:\basetype}
{\red{\int_{t_1}^{t_2}inl/inr~t~dx}{inl/inr~\int_{t_1}^{t_2}t~dx}}
\end{minipage}
\\
~\\
\begin{minipage}{0.90\hsize}
\infrule[EAppInt3]
{t_1,t_2:\basetype  }
{\red{\int_{t_1}^{t_2}\lambda y:T_2 .tdx}
     {\lambda y:T_2.\int_{t_1}^{t_2}t dx}}
\end{minipage}
\\
~\\
\begin{minipage}{0.90\hsize}
\infrule[EAppInt4]
{\forall i\in[1,n], t_{i*} = (t_{21}...,t_{2i-1},x_i,t_{1i+1}...,t_{1n})  }
{\red{\int_{(t_{11},t_{12},...t_{1n}) }^{(t_{21},t_{22},...,t_{2n})}tdx}
     {\int_{t_{11}}^{t_{21}}  \pi_1(t[t_{1*}/x])dx_1 \oplus   ... \oplus \int_{t_{1n}}^{t_{2n}} \pi_n(t[t_{n*}/x])dx_{n}}}
\end{minipage}

\caption{Reduction Rules for Derivative and Integration}
\label{fig:eval1}
\end{figure}

Our calculus is an extension of simply-typed \xmodify{lambda calculus}{lambda-calculus}. Our lambda abstraction and application are nothing different from the simply-typed lambda calculus, and we have the reduction rule:
\[
\red{(\lambda x:T.\,t) t_1}{t[t_1/x]}.
\]
%
%
We use an $n$-tuple to model structured data and projection $\pi_j$ to extract $j$-th component from a tuple, and we have the following reduction rule:
\[
\red{\pi_j (t_1,t_2,...t_{n})}{t_j}.
\]
Similarly, we have reduction rules for the case analysis:
\[
\reductiont
\]
\[
\reductionu
\]
Besides, we introduce fix-point operator to deal with recursion:
\[
\red{\mathit{fix}~f}{f~(\mathit{fix}~f)}
\]
It is worth noting that tuples, having a good correspondence in mathematics, should  be understood as structured data instead of high-dimensional vectors because there are some operations that are different from those in mathematics. As will be seen later, there is difference between our multiplication and matrix multiplication, and derivative and integration on tuples of tuples has no correspondence to mathematical objects.

The core reduction rules in our calculus are summarized in Figure \ref{fig:eval1}, which  define three basic cases for both reducing derivative terms and integration terms. For derivative, we use $\myder{t}{x}{t_0}$ to denote the derivative of $t$ over $x$ at point $t_0$, and we have four reduction rules:

\begin{itemize}
\item Rule {\sc EAppDer1} is to distribute point $t_0 : \basetype$ into a tuple.  This resembles the case in mathematics; if we have a function $f$ defined by  $f(x) = (f_1(x),f_2(x),\ldots,f_m(x))^T$, its derivative will be  $(\frac{\partial f_1}{\partial x}, \frac{\partial f_2}{\partial x},\ldots,\frac{\partial f_m}{\partial x})^T$.
For example, if we have a function $f : \mathbb{R} \to (\mathbb{R}, \mathbb{R})$ defined by $f(x) = (x,x*x)$, then its derivative will be $(1,2*x)$.
\item Rule {\sc EAppDer2} is similar to Rule {\sc EAppDer1}.
\item Rule {\sc EAppDer3} is to distribute point $t_0 : \basetype$ into a lambda abstraction. Again this is very natural in mathematics.
For example, for function $f(x) = \lambda y:B.\, x*y$, then we would have its derivative on $x$ as $\lambda y:B.y$.

\item Rule {\sc EAppDer4} is to deal with partial differentiation, similar to the Jacobian matrix in mathematics (as shown in the introduction).
For example, if we have a function that maps a pair $(x,y)$ to $(x*x,x*y \oplus y)$, which may be written as $\lambda z:(\basetype,\basetype).\,(\pi_1z*\pi_1z,(\pi_1z*\pi_2z\oplus\pi_2z))$  then we would have its derivative $\myder{(f~z)}{z}{(x,y)}$ as $((2*x,y),(0,x \oplus 1))$.

\end{itemize}

Similarly, we can define four reduction rules for integration. Rules {\sc EAppInt1},{\sc EAppInt2} and {\sc EAppInt3} are simple. Rule {\sc EAppInt4} is worth more explanation.
It is designed to establish the Newton-Leibniz formula
\[
\int_{t_1}^{t_2} \myder{t}{y}{x} dx = t[t_2/y] \ominus t[t_1/y]
\]
when $t_1$ and $t_2$ are tuples:
\[
\int_{(t_{11},t_{12},...,t_{1n})}^{(t_{21},t_{22},...,t_{2n})} \myder{t}{y}{x} dx
 = t[(t_{21},t_{22},...,t_{2n})/y]\ominus t[(t_{11},t_{12},...,t_{1n})/y].
\]
So we design the rule to have
\[
\int_{t_{1j}}^{t_{2j}}\myder{t[(t_{21},...,t_{2(j-1)},x_j',t_{1(j+1)},...,t_{1n})/y]}{x_j'}{x_j}dx_j
=
\int_{(t_{21},...,t_{2(j-1)},t_{1j},t_{1(j+1)} ,...,t_{1n}) }^{(t_{21},...,t_{2(j-1)},t_{2j},t_{1(j+1)} ,...,t_{1n})} \myder{t}{y}{x} dx.
\]
Notice that under our evaluation rules on derivative, $\pi_j(\frac{\partial t}{\partial x}|_{x=(x_1,x_2,...,x_{n})})$ will be equal to the derivative of $t$ to its $j$-th parameter $x_j$, so the integration will lead us to the original $t$.

\begin{figure}[t]
\centering
\ignore{
\begin{minipage}{0.80\hsize}
\infax[EAppAbs]
  {\red{(\lambda x:T.\,t)~v}{t[v/x]}}
\end{minipage}
\\
~\\
\begin{minipage}{0.80\hsize}
\infax[EAppProj]
{\red{\pi_j(t_1,t_2,...,t_{n})}{t_j}}
\end{minipage}
\\
~\\
}
\begin{minipage}{0.90\hsize}
\infax[EAppAdd1]
{\red{(t_{11},...,t_{1n}) \oplus (t_{21},...,t_{2n})}{(t_{11}\oplus t_{21},...,t_{1n}\oplus t_{2n})}}
\end{minipage}
\\
~\\
\begin{minipage}{0.90\hsize}
\infax[EAppAdd2]
{\red{(\lambda x:T.\,t_1) \oplus (\lambda y:T.\,t_2)}
     {\lambda x:T.\, (t_1 \oplus t_2[y/x])}}
\end{minipage}
\\
~\\
\begin{minipage}{0.90\hsize}
\infax[EAppSub1]
{\red{(t_{11},...,t_{1n}) \ominus (t_{21},...,t_{2n})}{(t_{11}\ominus t_{21},...,t_{1n}\ominus t_{2n})}}
\end{minipage}
\\
~\\
\begin{minipage}{0.90\hsize}
\infax[EAppSub2]
{\red{(\lambda x:T.\,t_1) \ominus (\lambda y:T.\,t_2)}
     {\lambda x:T.\, (t_1 \ominus t_2[y/x])}}
\end{minipage}
\\
~\\
\begin{minipage}{0.90\hsize}
\infrule[EAppMul1]
{t_0 :\basetype}
{\red{ (t_1,t_2,...,t_{n}) * t_0}
     {(t_1*t_0,t_2*t_0,...,t_{n}*t_0)}}
\end{minipage}
\\
~\\
\begin{minipage}{0.90\hsize}
\infrule[EAppMul2]
{t_0 :\basetype}
{\red{(\lambda x:T .t) * t_0}
     {\lambda x:T .(t * t_0)}}
\end{minipage}
\\
~\\
\begin{minipage}{0.90\hsize}
\infrule[EAppMul3]
{t_0 :\basetype}
{\red{(inl/inr~t) * t_0}
     {inl/inr~(t * t_0)}}
\end{minipage}
\\
~\\
\begin{minipage}{0.90\hsize}
\infrule[EAppMul4]
{t_1:(t_{11},t_{12},...t_{1n}),t_2 :(t_{21},t_{22},...t_{2n})}
{\red{t_1 * t_2}
     { (t_{11}*t_{21})\oplus (t_{12}*t_{22})\oplus ... \oplus (t_{1n}*t_{2n})}}
\end{minipage}
\caption{Reduction Rules for Addition, Subtraction and Multiplication}
\label{fig:eval2}
\end{figure}

Finally, we discuss the reduction rules for the three new binary operations, as summarized in Figure \ref{fig:eval2}.
The addition $\oplus$ is introduced to support the reduction rule of integration. It is also useful in proving the theorem and constructing the formula.
We can understand the two reduction rules for addition as the addition of high-dimension vectors and functions respectively.
Similarly, we can have two reduction rules for subtraction $\ominus$.
The operator $*$ was introduced as a powerful tool for constructing the Chain Rule and the Taylor's theorem.
The first two reduction rules can be understood as multiplications of a scalar with a function and a high-dimension vector respectively, while the last one can be understood as the multiplication on matrix. For example, we have
\[
((1,4),(2,5),(3,6))*(7,8,9)=(50,122)
\]
which corresponds to the following matrix multiplication.
\[
\begin{pmatrix}
  1 & 2 & 3\\
  4 & 5 & 6
\end{pmatrix}
\begin{pmatrix}
  7 \\
  8 \\
  9 \\
\end{pmatrix}
= \begin{pmatrix}
  50 \\ 122
\end{pmatrix}
\]
It is worth noting that while they are similar, $*$ \xmodify{is much different}{is different} from the matrix multiplication operation. For example, we cannot write $x$ as an $m$-dimensional vector (or $m*1$ matrix) in Taylor's theorem because no matrix $A$ is well-performed under $A*x*x$, but we can write Taylor's Theorem easily under our framework. In the matrix representation, the number of rows of the first matrix and the number of columns of the second matrix must be equal so that we can perform multiplication on them. This means, we can only write case $m=1$'s Taylor's theorem in matrices, while our version can perform for any tuples.

\subsection{Normal Forms}

\begin{figure}[thb]
\[
\begin{array}{lrcll}
\mbox{Normal Form} & \nf &::=& \nb & \mynote{normal form on $\iB$}\\
  && | & (\nf,\nf,\ldots,\nf) & \mynote{tuple} \\
  && | & \lambda x:T.\,t & \mynote{function, $t$ cannot be further reduced}\\
  && | & inl/inr~\nf & \mynote{injection}\\

\\
\mbox{Normal Forms on $\iB$} & \nb &::=& c & \mynote{constants on $\iB$}\\
 && | & x      &\mynote{variables on $\iB$} \\
 && | & \nb~\nf &\mynote{primitive function application} \\
 && | & \nb \oplus  \nf~~|~\nf \oplus  \nb & \mynote{addition}\\
 && | & \nb \ominus \nf~~|~\nf \ominus \nb & \mynote{subtraction} \\
 && | & \nb * \nb      &\mynote{multiplication} \\\vspace{1ex}
 && | & \myder{\nb}{x}{\nb} &\mynote{derivative}\xmodify{fix the space}{}\\
 && | & \myint{\nb}{\nb}{\nb} &\mynote{integration}\\
\end{array}
\]
\caption{Normal Forms}
\label{fig:normalForm}
\end{figure}

In our calculus, base type stands in a very special position, and we may involve many evaluations under the context of some free variables of an interpretable type. So for simplicity, we \xmodify{would}{will} use full reduction\footnote{By full reduction, we mean that a term can be reduced wherever any of its subterms can be reduced by a reduction rule.} but allow free variables of interpretable types (i.e., $\iB$) in our normal form.
Figure \ref{fig:normalForm} defines our normal form. It basically consists \xmodify{of normal form}{of the normal forms} on interpretable types, the tuple normal form, and the function normal form.

We have an interesting result about \xmodify{about normal form}{about the normal form} of a term of interpretable terms.

\begin{lemma}[Interpretability]
  \rm
All the normal forms of terms of interpretable types are interpretable terms.
That is, given a term $t : \iB$, if $t$ is in normal form, then $t$ is an interpretable term.
\end{lemma}
\begin{proof}
 We prove that a normal form $t$ is interpretable by induction on the form of $t$.
\begin{itemize}
\item Case $\lambda x:T.t$.
Because $\lambda x:T.t$ is of type $\iB$, $T$ must be of type $\iB$. Notice that the function body $t$ has a free variable $x$ of type $\iB$. By induction, we know $t$ is an interpretable term, therefore, $\lambda x:T.t$ is interpretable.

\item Case $(\nf,\nf,...,\nf)$.
This case is impossible, because it is not of type $\iB$. Using the same technique, we can prove the cases for $inl/inr~\nf$.

\item Case $c$. It is an interpretable term itself.

\item Case $\myder{\nb_1}{x}{\nb_2}$.
By induction, we have both $\nb_1$ and $\nb_2$ are interpretable terms. By definition of interpretable term $\itm$, we have $\myder{\itm}{x}{\itm}$ is an interpretable term. Thus $\myder{\nb_1}{x}{\nb_2}$ is an interpretable term. Using the same technique, we can prove the cases for $\nb*\nb$, $\myder{\nb}{x}{\nb}$, and $\myint{\nb}{\nb}{\nb}$.

\item Case $\nb~\nf$.
$\nb$ is of type $\iB$, and $\nb~\nf$ is of type $\iB$. Thus we can deduce that $\nf$ is of type $\iB$. By induction both $\nb$ and $\nf$ are interpretable terms. Thus $\nb~\nf$ is an interpretable term. Using the same technique, we can prove the cases for $\nb\oplus \nf$, $\nf\oplus \nb$, $\nb \ominus \nf$, and $\nf \ominus \nb$.
\end{itemize}
\end{proof}

\subsection{Properties}

Next, we prove some properties of our calculus. The proof is rather routine with some small variations.

\ignore{
Before we enter the proof of soundness, we first take a brief look at how our calculus works, and get some insight into why these properties hold in our calculus.

The main difference between our calculus and normal lambda calculus is that it requires evaluation under the context of several free variables of Interpretable Type.
As an example, consider $\myder{t_1}{x}{t_2}$ where $t_2$ is a term of base type.
As the evaluation rules that can be applied to this term are only Rules ({\sc EAppDer1}) and ({\sc EAppDer2}),
one may wonder how to deal with the term $\myder{(t_3\oplus t_4)}{x}{t_2}$. The answer is that under a full reduction strategy, and later we will prove we can always reduce $t_3\oplus t_4$ to a form such that the above rules can be applied.

And besides the case of there are only free variables of Interpretable Types, we may encounter situations such as there is a free variable y of type (\basetype,\basetype), in the context. But in this case, we will find the variable y is introduced by $\lambda y:(B,B).\, t$, $\myder{t}{y}{t}$ or $\int_t^t t dy$. In the first case, the lambda abstraction is a normal form itself, which means we do not need to conduct reductions on its function body. In the second and third case, we have the evaluation rules

\reductiong

\reductionh
to erase the occurrence of variable $y$. And finally, we would come to a context that we conduct reduction only with free variables of Interpretable Types.

}
\begin{lemma}[Properties]
  \rm
This calculus has the properties of progress, preservation and confluence. Moreover, if a term $t$ does not contain subterms $fix~t'$, then $t$ is strong normalizable.
\end{lemma}

\begin{proof}
Full proof is in the \xmodify{appendix}{Appendix [\ref{ProgressProof},\ref{PreservationProof},\ref{ConfluenceProof},\ref{NormalizationProof}]}, which is adapted from the standard proof.
\end{proof}

\subsection{Term Equality}

We need to talk a bit more on equality because we do not consider reduction or calculation on primitive functions. This notion of equality has little to do with our evaluation but has a lot to do with the equality of primitive functions. Using this notion of equality, we can compute the result from completely different calculations. This will be used in our later proof of the three theorems.

Since we have proved the confluence property, we know that every term has at most one normal form after reduction. Thus, we can define our equality based on their normal forms; the equality between unnormalizable terms is undefined.

%

\xmodify{I change it into a definition}{}
\begin{definition}[Term Equality]
\rm
  An open term $t_1$ is said to be equal to a term $t_2$, if and only if for all free variables $x_1,x_2,...,x_n$ in $t_1$ and $t_2$, \xmodify{$t_1[u_1/x_1,..., u_n/x_n]=t_2[u_1/x_1,..., u_n/x_n]$, where $u_i$ is a closed and normalizable term (Here normalizable means it has a normal form).}{for all closed and weak-normalizable term $u_i$ whose type is the same as that of $x_i$, we have $t_1[u_1/x_1,..., u_n/x_n]=t_2[u_1/x_1,..., u_n/x_n]$.}

A closed-term $t_1 = t_2$, if their normal forms $n_1$ and $n_2$ have the relation that $n_1 = n_2 $, where
a normal form $n_1$ is said to be equal to another normal form $n_2$, if they satisfy one of the following rules:
\begin{itemize}
\item(1)
$n_1$ is a of type $\iB$, then $n_2$ has to be of the same type, and under the base type interpretation, $n_1$ is equal to $n_2$;
\item(2)
$n_1$ is $(t_1,t_2,...,t_n)$, then $n_2$ has to be $(t_1',t_2',...,t_n')$, and $\forall j\in[1,n],t_j$ is equal to $t_j'$;
\item(3)
$n_1$ is  $\lambda x:T. t$, then $n_2$ has to be $\lambda y :T. t'$ ($y$ can be $x$), and  $n_1~x$ is equal to $n_2~x$.
\item(4)
$n_1$ is  $inl~t_1'$, then $n_2$ has to be $inl~t_2'$, and $t_1'$ is equal to $t_2'$.
\item(5)
$n_1$ is  $inr~t_1'$, then $n_2$ has to be $inr~t_2'$, and $t_1'$ is equal to $t_2'$.

\end{itemize}
\end{definition}

\begin{lemma}
  \rm
 The equality is reflexive, transitive and symmetric for weak-normalizable terms.
\end{lemma}

\begin{proof}
 Based on the equality of terms of base types, we can prove it by induction.
\end{proof}

\begin{lemma}
  \rm
 The equality is consistent, e.g., we can not prove equality between arbitrary two terms.
\end{lemma}

\begin{proof}
Notice that except for the equality introduced by the base type interpreter, other equality \xmodify{inference}{inferences} all preserve the type. So for arbitrary $t_1$ of type $(B,B)$ and $t_2$ of type $B$, we can not prove equality between them.
\end{proof}

Next we give some lemmas that will be used later in our proof. It is relatively unimportant to the mainline of our calculus, so we put their proofs in the Appendix.

\begin{lemma}\label{ReSub}
  \rm
\xmodify{if}{If} $t_1 \rho^* t_1',t_2 \rho^* t_2'$,~\xmodify{none}{then}~$t_1[t_2/x] \rho^* t_1'[t_2'/x]$.\end{lemma}


\begin{lemma}\label{EqAdd}
\rm

\xmodify{if}{If}  $t_1 =t_1'$, $t_2=t_2'$, then $t_1\oplus t_2 = t_1'\oplus t_2' $.
\end{lemma}

\begin{lemma}\label{EqLSub}
\rm
For a term t, for any subterm s, if the term s'=s, then t[s'/s]=t. (We only substitute the subterm s, but not other subterms same as s)
\end{lemma}

\begin{lemma}\label{EqDist}
\rm
If $t_1*(t_2\oplus t_3)$ and $(t_1*t_2)\oplus (t_1*t_3)$ are weak-normalizable, then $t_1*(t_2\oplus t_3) = (t_1*t_2)\oplus (t_1*t_3)$.
\end{lemma}

\begin{lemma}\label{EqCom}
\rm
If $(t_1\ominus t_2)\oplus(t_2 \ominus t_3)$ and $t_1\ominus t_3$ are weak-normalizable, then $(t_1\ominus t_2)\oplus(t_2 \ominus t_3) = t_1\ominus t_3$.
\end{lemma}


\section{Newton-Leibniz's Formula}
\label{sec:NL}

The first important theorem we will give is the Newton-Leibniz's formula, which ensures the duality between derivatives and integration. This theorem lays a solid basis for our calculus. Before giving and proving the theorem, as a warmup, let us take a look at a simple calculation example related to derivative and integration.

\begin{example}[Calculation with Derivatives and Integrations] Consider a function $f$ on real numbers, usually defined in mathematics as $f (x,y) = (x+y, x*y, y)$. In our calculus, it is defined as follows.
\[
\begin{array}{llll}
 \m{f} &::& (\mathbb{R},\mathbb{R})\to (\mathbb{R},\mathbb{R},\mathbb{R}) \\
 \m{f}  &=& \lambda x:(\mathbb{R},\mathbb{R}).\,(\pi_1(x)\oplus \pi_2(x),\pi_1(x)*\pi_2(x),\pi_2(x))\\
\end{array}
\]

The following shows the calculation of how $\int_{(0,0)}^{(2,3)}\myder{(f\,y)}{ y}{x}dx$ comes equal with $f~y[(2,3)/y]\ominus f~y[(0,0)/y]$
\[
\begin{array}{llll}
  & & \int_{(0,0)}^{(2,3)}\myder{(f\,y)}{ y}{x}dx\\
  &=& \reason{Rule {\sc EAppInt3}}\\
  & & \int_0^2\pi_1(\myder{(f\,y)}{ y}{(x_1,0)})dx_1\oplus \int_0^3\pi_2(\myder{(f\,y)}{ y}{(2,x_2)})dx_2\\

&=& \reason{Rule {\sc EAppDer3} }\\
& & \int_0^2\pi_1(\myder{ f (x_1',0)}{ x_1'}{x_1},\myder{ f (x_1,x_2')}{ x_2'}{x_2})dx_1\oplus \int_0^3\pi_2(\myder{ f (x_1',x_2)}{ x_1'}{x_1},\myder{ f (2,x_2')}{ x_2'}{x_2})dx_2\\

&=& \reason{Projection }\\
& & \int_0^2\myder{ f (x_1',0)}{ x_1'}{x_1}dx_1\oplus \int_0^3\myder{ f (2,x_2')}{ x_2'}{x_2}dx_2\\

&=& \reason{Function Application}\\
& & \int_0^2\myder{ (x_1'\oplus 0,x_1'*0,0)}{ x_1'}{x_1}dx_1\oplus \int_0^3\myder{ (2\oplus x_2',2*x_2',x_2')}{ x_2'}{x_2}dx_2\\

\end{array}
\]

\[
\begin{array}{llll}

&=& \reason{Rule {\sc EAppInt1}}\\
& & \int_0^2(\myder{ x_1'\oplus 0}{ x_1'}{x_1},\myder{ x_1'*0}{ x_1'}{x_1},\myder{0}{ x_1'}{x_1})dx_1\oplus \int_0^3(\myder{ 2\oplus x_2'}{ x_2'}{x_2},\myder{ 2*x_2'}{ x_2'}{x_2},\myder{  x_2'}{ x_2'}{x_2})dx_2\\

&=& \reason{Lemma \ref{EqLSub}} \\
& & \int_0^2(1,0,0)dx_1\oplus \int_0^3(1,2,1)dx_2\\

&=&\reason{Rule {\sc EAppInt1}}\\
& & (2,0,0)\oplus (3,6,3)\\

&=&\reason{Rule {\sc EAppAdd1}}\\
& & (5,6,3)
\end{array}
\]

For readability, we substitute the subterm $(\myder{ x_1'\oplus 0}{ x_1'}{x_1},\myder{ x_1'*0}{ x_1'}{x_1},\myder{0}{ x_1'}{x_1})$ and $(\myder{ 2\oplus x_2'}{ x_2'}{x_2},\myder{ 2*x_2'}{ x_2'}{x_2},\myder{  x_2'}{ x_2'}{x_2})$ for the same subterm $(1,0,0)$ and $(1,2,1)$ during the calculation, though our calculus does not actually perform computation like this. These substitutions are safe to perform (Lemma \ref{EqLSub}), and give a better demonstration on how the Newton-Leibniz theorem works. And Here we have $\int_{(0,0)}^{(2,3)}\myder{(f\,y)}{ y}{x}dx$ $=(5,6,3)$ $=f(2,3)\ominus f(0,0)$ $=f~y[(2,3)/y]\ominus f~y[(0,0)/y]$.

\myend
\end{example}

\begin{theorem}[Newton-Leibniz]
\label{theorem:nl}
\rm
 Let t contain no free occurrence of $x$, and both $\int_{t_1}^{t_2} \myder{ t}{ y}{x} dx$ and $t[t_2/y]\ominus t[t_1/y]$ are well-typed and weak-normalizable. Then we have
 \[
 \int_{t_1}^{t_2} \myder{ t}{ y}{x} dx = t[t_2/y]\ominus t[t_1/y].
 \]
\end{theorem}

\begin{proof}

If $t_1$, $t_2$ or $t$ is not closed, then we need to prove $\forall u_1,...,u_n$, we have \[
(\int_{t_1}^{t_2} \myder{ t}{ y}{x} dx)[u_1/x_1,..., u_n/x_n] = (t[t_2/y]\ominus t[t_1/y])[u_1/x_1,..., u_n/x_n].
\]
By freezing $u_1,...,u_n$, we can apply the substitution $[u_1/x_1,..., u_n/x_n]$ to make every term closed. So, for simplicity, we will assume $t$, $t_1$ and $t_2$ to be closed.

We prove this by induction on types.

\begin{itemize}

\item Case: $t_1$,$t_2$ and $t$ are of base types. By the confluence lemma, we know there exists the normal form $t'$, $t_1'$ and $t_2'$ of the term $t$, $t_1$ and $t_2$. Also, we know $\int_{t_1}^{t_2} \myder{ t}{ y}{x} dx = \int_{t_1'}^{t_2'} \myder{ t'}{ y}{x} dx$ and $t[t_2/y]\ominus t[t_1/y] = t'[t_2'/y]\ominus t'[t_1'/y]$.
Since on base types we have $\int_{t_1'}^{t_2'} \myder{ t'}{ y}{x} dx = t'[t_2'/y]\ominus t'[t_1'/y]$, we have $\int_{t_1}^{t_2} \myder{ t}{ y}{x} dx = t[t_2/y]\ominus t[t_1/y]$.

\smallskip

\item Case: $t_1$,$t_2$ are of base types, $t$ is of type $(T_1,T_2,...,T_{n})$. By the
confluence lemmas, there exist a normal form $(t_{11}',t_{12}',...,t_{1n}')$ for $t$. Using Rules ({\sc EAppInt1}) and ({\sc EAppDer1}),
we know
\[
\begin{array}{llll}
  \int_{t_1}^{t_2} \myder{ t}{ y}{x} dx &=& \int_{t_1}^{t_2} \myder{ (t_{11}',t_{12}',...,t_{1n}')}{ y}{x} dx\\
 &=&  \int_{t_1}^{t_2}  (\myder{ t_{11}'}{ y}{x},\myder{ t_{12}'}{ y}{x},...,\myder{ t_{1n}'}{ y}{x}) dx\\
 &=& (\int_{t_1}^{t_2}\myder{ t_{11}'}{ y}{x}dx,
\int_{t_1}^{t_2}\myder{ t_{12}'}{ y}{x}dx,...,\int_{t_1}^{t_2}\myder{ t_{1n}'}{ y}{x}dx)
\end{array}
\]
On the other hand, we have
\[
\begin{array}{llll}
t[t_2/y]\ominus t[t_1/y]\\
\qquad = (t_{11}'[t_2/y],t_{12}'[t_2/y],...,t_{1n}'[t_2/y])\ominus (t_{11}'[t_1/y],t_{12}'[t_1/y],...,t_{1n}'[t_1/y])\\
\qquad = (t_{11}'[t_2/y]\ominus t_{11}'[t_1/y],t_{12}'[t_2/y]\ominus t_{12}'[t_1/y],...,t_{1n}'[t_2/y]\ominus t_{1n}'[t_1/y])
\end{array}
\]
By induction, we have $\forall j\in[1,n],\int_{t_1}^{t_2}\myder{ t_{1j}'}{ y}{x}dx = t_{1j}'[t_2/y]\ominus t_{1j}'[t_1/y]$ \xmodify{, so we prove}{, so we have proven} the case.


\smallskip

\item Case: $t_1$,$t_2$ are of base types, $t$ is of type $A\rightarrow B$.
By Lemma \ref{EqLSub}, we can use $\lambda z:A. t~z$ (for simiplicity, we use $\lambda z:A. t'$ where $t'$ = $t~z$) to substitute for $t$, where z is a fresh variable. Now, we have for any $u$,
\[
\begin{array}{llll}
  (\int_{t_1}^{t_2} \myder{ t}{ y}{x} dx)~ u
  &=& (\int_{t_1}^{t_2} \myder{ \lambda z:A .t'}{ y}{x} dx)~ u \\
  &=& \lambda z:A .(\int_{t_1}^{t_2} \myder{ t'}{ y}{x} dx)~ u \\
  &=& \int_{t_1}^{t_2} \myder{ t'[u/z]}{ y}{x} dx
\end{array}
\]
and on the other hand, since $z$ is free in $t_1$ and $t_2$,  we have
\[
\begin{array}{llll}
  (t[t_2/y]\ominus t[t_1/y])~u
  &=& ((\lambda z:A.t')[t_1/y]\ominus(\lambda z:A.t')[t_2/y])~u\\
  &=& \lambda z:A.(t'[t_2/y]\ominus t'[t_1/y])~u \\
  &=& (t'[t_2/y]\ominus t'[t_1/y])[u/z] \\
  &=& (t'[u/z])[t_2/y]\ominus (t'[u/z])[t_1/y]
\end{array}
\]

By induction (on $B$), we know $\int_{t_1}^{t_2} \myder{ t'[u/z]}{ y}{x} dx = (t'[u/z])[t_2/y]\ominus (t'[u/z])[t_1/y]$\xmodify{, thus we prove}{, thus we have proven} the case.

\smallskip

\item Case: $t_1$,$t_2$ are of base types, $t$ is of type $T_1+T_2$.
This case is impossible because the righthand term is not well-typed.

\item Case: $t_1$,$t_2$ are of type $(T_1,T_2,...,T_{n})$, $t$ is of any type $T$. By using the confluence lemma, we know there exist the normal forms $(t_{11}',t_{12}',...,t_{1n}')$ and $(t_{21}',t_{22}',...,t_{2n}')$ for $t_1$ and $t_2$ respectively.

Applying Rules (EAppDer3) and (EAppInt3), we have
\[
\begin{array}{llll}
\int_{t_1}^{t_2} \myder{ t}{ y}{x} dx
 &=& \int_{(t_{11}',t_{12}',...,t_{1n}')}^{(t_{21}',t_{22}',...,t_{2n}')} \myder{ t}{ y}{x} dx \\
 &=& \int_{t_{11}'}^{t_{21}'}  \pi_1(\myder{ t}{ y}{x}[(x_1,t_{12}',...t_{1n}')/x])dx_1 \oplus  \cdots \oplus \\
 & & \quad  \int_{t_{1n}'}^{t_{2n}'} \pi_n(\myder{ t}{ y}{x}[(t_{21}',t_{22}',...,x_{n})/x])dx_{n}
\end{array}
\]
Notice that there is no occurrence of $x$ in $t$, so we have
\[
\begin{array}{llll}
\int_{t_{1j}'}^{t_{2j}'}  \pi_j(\myder{ t}{ y}{x}[(t_{21}',t_{22}',...,t_{2(j-1)}',x_j,t_{1(j+1)}',...,t_{1n}')/x])dx_j\\
\qquad = \int_{t_{1j}'}^{t_{2j}'}  \pi_j(\myder{ t}{ y}{(t_{21}',t_{22}',...,t_{2(j-1)}',x_j,t_{1(j+1)}',...,t_{1n}')})dx_j\\
\qquad = \int_{t_{1j}'}^{t_{2j}'}  \pi_j (\myder{ t[t_{1*}/y]}{ x_1}{t_{21}'},\myder{ t[t_{2*}/y]}{ x_2}{t_{22}'},...,\myder{ t[t_{(j-1)*}/y]}{ x_{j-1}}{t_{2(j-1)}'},\\
\qquad\qquad\qquad \myder{ t[t_{j*}/y]}{ x_j'}{x_j},\myder{ t[t_{(j+1)*}/y]}{ x_{j+1}}{t_{1(j+1)}'},...,\myder{ t[t_{n*}/y]}{ x_{n}}{t_{1n}'})dx_j\\
\qquad = \int_{t_{1j}'}^{t_{2j}'}\myder{ t[(t_{21}',t_{22}',...,t_{2(j-1)}',x_j',t_{1(j+1)}',...,t_{1n}')/y]}{ x_j'}{x_j}dx_j
\end{array}
\]
By induction (on the case where $t_1$, $t_2$ are of type $T_j$, $t$ is of type $T$), we have
\[
\begin{array}{llll}
\int_{t_{1j}'}^{t_{2j}'}\myder{ t[(t_{21}',t_{22}',...,t_{2(j-1)}',x_j',t_{1(j+1)}',...,t_{1n}')/y]}{ x_j'}{x_j}dx_j\\
\qquad = (t[(t_{21}',t_{22}',...,t_{2(j-1)}',x_j',t_{1(j+1)}',...,t_{1n}')/y])[t_{2j}'/x_j'] ~\ominus \\
\qquad\qquad\qquad (t[(t_{21}',t_{22}',...,t_{2(j-1)}',x_j',t_{1(j+1)}',...,t_{1n}')/y])[t_{1j}'/x_j']\\
\qquad = (t[(t_{21}',t_{22}',...,t_{2(j-1)}',t_{2j}',t_{1(j+1)}',...,t_{1n}')/y]) ~\ominus\\
\qquad\qquad\qquad (t[(t_{21}',t_{22}',...,t_{2(j-1)}',t_{1j}',t_{1(j+1)}',...,t_{1n}')/y])
\end{array}
\]
Note that the last equation holds because $x_j'$ is a fresh variable and $t$ has no occurrence of $x_j'$.

Now we have the following calculation.
\[
\begin{array}{llll}
& & \int_{t_1}^{t_2} \myder{ t}{ y}{x} dx\\
&=&  \reason{all the above}\\
& & ((t[(t_{21}',t_{12}',...,t_{1n}')/y]) \ominus (t[(t_{11}',t_{12}',...,t_{1n}')/y])) ~\oplus\\
& & \qquad ((t[(t_{21}',t_{22}',...,t_{1n}')/y]) \ominus (t[(t_{21}',t_{12}',...,t_{1n}')/y])) \oplus \cdots \oplus\\
& & \qquad \qquad ((t[(t_{21}',t_{22}',...,t_{2n}')/y]) \ominus (t[(t_{21}',t_{22}',...,t_{1n}')/y]))\\
 &=& \reason{Lemma \ref{EqCom}}\\
 & &  (t[(t_{21}',t_{22}',...,t_{2n}')/y]) \ominus (t[(t_{11}',t_{12}',...,t_{1n}')/y])\\
 &=& \reason{Lemma \ref{ReSub}}\\
 & & t[t_2/y]\ominus t[t_1/y]
\end{array}
\]
\end{itemize}

\xmodify{none}{Thus we have proven the theorem.}
\end{proof}

\subsection*{Application: Incremental Computation}
A direct application is incrementalization \cite{Liu00,CGRO14,GiGS19}. Given a function $f(x)$, if the input $x$ is changed by $\Delta$, then we can obtain its incremental version of $f(x)$ by $f'(x,\Delta)$.
\[
f(x \oplus \Delta) = f(x) \oplus f'(x,\Delta)
\]
where $f'$ satisfies that
\[
f'(x,\Delta) = \myint{x}{x \oplus \Delta}{\myder{f(x)}{x}{x}}.
\]


\begin{example}[Averaging a Pair of Real numbers]
As a simple example, consider the average of a pair of real numbers
\[
\begin{array}{llll}
 \m{average} &::& (\mathbb{R},\mathbb{R}) \to \mathbb{R}\\
 \m{average}~ &=& \lambda x. (\pi_1(x) + \pi_2(x))/2\\
 \end{array}
\]
Suppose that we want to get an incremental computation of $\m{average}$ at $x = (x_1,x_2)$ when the first element $x_1$ is changed to $x_1 + d$ while the second component $x_2$ is kept the same. The incremental computation is defined by
\[
\m{inc}(x,d) = \m{average}(x,(d,0)) = \myint{x}{x \oplus (d,0)}{\myder{\m{average}(x)}{x}{x}} = \frac{d}{2}
\]
which is efficient.
\end{example}
\ignore{
Suppose there is a bag of integers 1,2,3 and 4, we got an average of 2.5. And now we add changes (for example, add 2) for each integer, what is the average now?

From The Theorem 1 we know the incrementation can be computed by $\int_{t_1}^{t_2} \myder{ t}{ y}{x} dx = t[t_2/y]\ominus t[t_1/y]$,

Notice that for all Base Type $C_1$,$C_2$,$C_3$,$C_4$, we have
\[
\begin{array}{llll}
\myder{ \m{average}~xs}{ xs}{xs=(C_1,C_2,C_3,C_4)}\\
\qquad = (\myder{ \m{sum}~(x_1,C_2,C_3,C_4) / \m{length}~(x_1,C_2,C_3,C_4)}{ x_1}{C_1},\myder{ \m{sum}~(C_1,x_2,C_3,C_4) / \m{length}~(C_1,x_2,C_3,C_4)}{ x_2}{C_2},\\
\myder{ \m{sum}~(C_1,C_2,x_3,C_4) / \m{length}~(C_1,C_2,x_3,C_4)}{ x_3}{C_3},\myder{ \m{sum}~(C_1,C_2,C_3,x_4) / \m{length}~(C_1,C_2,C_3,x_4)}{ x_4}{C_4})\\
\qquad =(\frac{1}{4},\frac{1}{4},\frac{1}{4},\frac{1}{4})
\end{array}
\]
So the change would be
\[
\begin{array}{llll}
\int_{(1,2,3,4)}^{(3,4,5,6)} (\frac{1}{4},\frac{1}{4},\frac{1}{4},\frac{1}{4}) dx\\
=\int_{1}^{3}\frac{1}{4}dx_1\oplus\int_{2}^{4}\frac{1}{4}dx_1\oplus\int_{3}^{5}\frac{1}{4}dx_1\oplus\int_{4}^{6}\frac{1}{4}dx_1\\
=2
\end{array}
\]

So we got an average of 3,4,5,6 as 2+2.5=4.5. This computation presents the complete process of how to compute the change, which may seem a little bit complex. But we can make many optimizations here because it is easy to find that the derivative of the result is constant. The user can use the constant to programme the derivatives and integrate them into the change, just like the way the change theory does with the function Derive.
}

\section{Chain Rule}
\label{sec:chain}

The Chain Rule is another important theorem of the relation between function composition and derivatives. This Chain Rule in our calculus has many important applications in automatic differentiation and incremental computation. We first give an example to get some taste, before we give and prove the theorem.

\begin{example}[chain rule] Consider two functions $f$ and $g$ on real numbers, usually defined in mathematics as $f (x,y) = (x+y, x*y, y)$ and $g (x,y) = (x+y,y)$. In our calculus, they are defined as follows.
\[
\begin{array}{llll}

 \m{f} &::& (\mathbb{R},\mathbb{R})\to (\mathbb{R},\mathbb{R},\mathbb{R}) \\
 \m{f}  &=& \lambda x:(\mathbb{R},\mathbb{R}).(\pi_1(x)\oplus \pi_2(x),\pi_1(x)*\pi_2(x),\pi_2(x))\\
 \m{g} &::& (\mathbb{R},\mathbb{R})\to (\mathbb{R},\mathbb{R})\\
 \m{g}  &=& \lambda x:(\mathbb{R},\mathbb{R}). (\pi_1(x)\oplus \pi_2(x),\pi_2(x))\\
\end{array}
\]
We demostrate that for any $r_1,r_2,r_3,r_4 \in \mathbb{R}$,
we have
\[
\myder{ f(g~x)}{ x}{(r_3,r_4)}*(r_1,r_2) = \myder{ f\,x}{ x}{g~(r_3,r_4)}*(\myder{ g\,x}{ x}{(r_3,r_4)}*(r_1,r_2))
\]
by the following calculation. First, for the LHS, we have:
\[
\begin{array}{llll}

& & \myder{ f (g~x)}{ x}{(r_3,r_4)}*(r_1,r_2)\\
&=& \reason{Rule {\sc EAppDer3}}\\

 &&\myder{ f (g~(x_1,r_4))}{ x_1}{r_3}*r_1\oplus \myder{ f (g~(r_3,x_2))}{ x_2}{r_4}*r_2\\
&=& \reason{Application}\\
&&\myder{ (x_1\oplus r_4\oplus r_4,(x_1\oplus r_4)*r_4,r_4)}{ x_1}{r_3}*r_1\oplus\myder{ (r_3\oplus x_2\oplus x_2,(r_3\oplus x_2)*x_2,x_2)}{ x_2}{r_4}*r_2\\

&=& \reason{Lemma \ref{EqLSub}}\\
&& (1,r_4,0)*r_1\oplus (2,r_3\oplus 2*r_4,1)*r_2\\

&=& \reason{Rule {\sc EAppMul1} and Rule {\sc EAppAdd1}}\\
&& (r_1\oplus(2*r_2),r_4*r_1\oplus(r_3\oplus (2*r_4))*r_2,r_2)\\

\end{array}
\]
Now, for the RHS, we calculate with the following two steps.
\[
\begin{array}{llll}

& & \myder{ g~x}{ x}{(r_3,r_4)}*(r_1,r_2)\\
&=& \reason{Rule {\sc EAppDer3}}\\

&&\myder{ g~(x_1,r_4)}{ x_1}{r_3}*r_1\oplus \myder{ g~(r_3,x_2)}{ x_2}{r_4}*r_2\\

&=&\reason{Application}\\
&&\myder{ (x_1\oplus r_4,r_4)}{ x_1}{r_3}*r_1\oplus \myder{ (r_3\oplus x_2,x_2)}{ x_2}{r_4}*r_2\\

&=&  \reason{Lemma \ref{EqLSub}}\\
&& (1,0)*r_1\oplus (1,1)*r_2\\

&=&  \reason{Rule {\sc EAppMul1} and Rule {\sc EAppAdd1}}\\
&&(r_1\oplus r_2,r_2)\\
\\
&&\myder{ f~x}{ x}{g~(r_3,r_4)}*(r_1\oplus r_2,r_2)\\

&=&\reason{Application and Rule {\sc EAppDer3}}\\

&&(\myder{ f~(x_1,r_4)}{ x_1}{r_3\oplus r_4},\myder{ f~(r_3\oplus r_4,x_2)}{ x_2}{r_4})*(r_1\oplus r_2,r_2)\\

&=&\reason{Application}\\

&&(\myder{ (x_1\oplus r_4,x_1*r_4,r_4)}{ x_1}{r_3\oplus r_4},\myder{ (r_3\oplus r_4\oplus x_2,(r_3\oplus r_4)*x_2,x_2)}{ x_2}{r_4})*(r_1\oplus r_2,r_2)\\

&=&  \reason{Lemma \ref{EqLSub}}\\
&& ((1,r_4,0),(1,(r_3\oplus r_4),1))*(r_1\oplus r_2,r_2)\\

&=&\reason{Rule {\sc EAppMul4}, Rule {\sc EAppAdd1} and Lemma \ref{EqLSub}}\\

&&(r_1\oplus(2*r_2),r_4*r_1\oplus(r_3\oplus (2*r_4))*r_2,r_2)\\

\end{array}
\]
\myend
\end{example}


\begin{theorem}[Chain Rule]
  \label{theorem:chainRule}
\rm
Let $f: T_1 \to T$, $g : T_2 \to T_1$. If both $
\myder{ f(g~x)}{ x}{t_1}*t$ and $\myder{ f\,y}{ y}{(g~t_1)}*(\myder{ g\,z}{ z}{t_1}*t)$ are well-typed and weak-normalizable. Then for any $t,t_1 : T_2$, we have
\[
\myder{ f(g~x)}{ x}{t_1}*t=\myder{ f\,y}{ y}{(g~t_1)}*(\myder{ g\,z}{ z}{t_1}*t).
\]
\end{theorem}

\begin{proof}

Like in the proof of Theorem \ref{theorem:nl}, for simplicity, we assume that $f$, $g$, $t$ and $t_1$ are closed. Furthermore, we assume that $t$ and $t_1$ are in normal form.
We prove this by induction on types.

\begin{itemize}

\item Case $T,T_2$ are base types, and $T_1$ is any type. To be well-typed, $T_1$ must contain no $\rightarrow$ or $+$ type. So for simplicity, we suppose $T_1$ to be $(B,B,B,...,B)$ of $n$-tuples, but the technique below can be applied to any $T_1$ type (such as tuples of tuples) that  makes the term well-typed.

First we notice that
\[
\begin{array}{llll}
g~z & =& (\pi_1(g~z),\pi_2(g~z),...,\pi_n(g~z))\\
& = &((\lambda b':B. \pi_1(g~b'))~z, (\lambda b':B. \pi_2(g~b'))~z,..., (\lambda b':B. \pi_n(g~b'))~z)
\end{array}
\]

and for any $j$, we notice that $\pi_j(g~b')$ has only one free variable of base type, so it can be reduced to a normal form, say $E_j$, of base type. Let $g_j$ be $\lambda b':B.E_j$, then we have $g~z = (g_1~z,g_2~z,...,g_n~z)$.

Next, we deal with the term \xmodify{f}{$f$}:
\[
\begin{array}{llll}
f &=&  \lambda a:T_1.\,(f~a)\\
 &=& \lambda a:T_1.\, ((\lambda y_1:B.\,\lambda y_2:B.,...\lambda y_{n}:B.\,(f~(y_1,y_2,...,y_n)))~\pi_1(a)~\pi_2(a)...~\pi_n(a))
\end{array}
\]
and we know that
$(f~(y_1,y_2,...,y_n))$ only contains base type free variables, so it can be reduced to a base type normal form, say $N$, so we have
\[
\begin{array}{llll}
f= \lambda a:T.\,((\lambda y_1:B.\lambda y_2:B.,...\lambda y_{n}:B.\,N)~\pi_1(a)~\pi_2(a)...~\pi_n(a)).
\end{array}
\]
Now, we can calculate as follows:
\[
\begin{array}{llll}
\myder{ f(g~x)}{ x}{t_1}*t\\
\qquad = \myder{ (\lambda a:T. (\lambda y_1:B.\lambda y_2:B.,...\lambda y_{n}:B.N)~\pi_1(a)~\pi_2(a)...~\pi_n(a))~(g_1~x,g_2~x,...,g_{n}~x)}{ x}{t_1}*t\\
\qquad = \myder{(\lambda y_1:B.\lambda y_2:B.,...\lambda y_{n}:B.N)~(g_1~x)~(g_2~x)...~(g_{n}~x)}{ x}{t_1}*t\\\vspace{1ex}
\qquad = \myder{ N[(g_1~x)/y_1,(g_2~x)/y_2,...(g_{n}~x)/y_{n}]}{ x}{t_1}*t\\
\\
\myder{ f~y}{ y}{(g~t_1)}*(\myder{ g~z}{ z}{t_1}*t) \\\vspace{1ex}
\qquad=  \myder{ f~y}{ y}{(g_1~t_1,g_2~t_1,...,g_{n}~t_1)}*(\myder{ (g_1~z,g_2~z,...,g_{n}~z)}{ z}{t_1}*t) \\\vspace{1ex}
\qquad= \myder{ f~y}{ y}{(g_1~t_1,g_2~t_1,...,g_{n}~t_1)}*(\myder{ g_1~z}{ z}{t_1}*t,\myder{ g_2~z}{ z}{t_1}*t,...,\myder{ g_{n}~z}{ z}{t_1}*t)\\\vspace{1ex}
\qquad=\myder{ (\lambda y_1:B.\lambda y_2:B.,...\lambda y_{n}:B.N)~\pi_1(y)~\pi_2(y)...~\pi_n(y)}{ y}{(g_1~t_1,g_2~t_1,...,g_{n}~t_1)}*\\\vspace{1ex}
\qquad\qquad\qquad(\myder{ g_1~z}{ z}{t_1}*t,\myder{ g_2~z}{ z}{t_1}*t,...,\myder{ g_{n}~z}{ z}{t_1}*t)\\\vspace{1ex}
\qquad= (\myder{ N[y_1'/y_1,g_2~t_1/y_2,...,g_{n}~t_1/y_{n}]}{ y_1'}{g_1~t_1},...,\myder{ N[g_1~t_1/y_1,g_2~t_1/y_2,...,y_{n}'/y_{n}]}{ y_{n}'}{g_{n}~t_1})*\\\vspace{1ex}
\qquad\qquad\qquad(\myder{ g_1~z}{ z}{t_1}*t,\myder{ g_2~z}{ z}{t_1}*t,...,\myder{ g_{n}~z}{ z}{t_1}*t)\\\vspace{1ex}
\qquad= (\myder{ N[y_1'/y_1,g_2~t_1/y_2,...,g_{n}~t_1/y_{n}]}{ y_1'}{g_1~t_1}*(\myder{ g_1~z}{ z}{t_1}*t))\oplus ... \oplus \\\vspace{1ex}
\qquad\qquad\qquad(\myder{  N[g_1~t_1/y_1,g_2~t_1/y_2,...,y_{n}'/y_{n}]}{ y_{n}'}{g_{n}~t_1}*(\myder{ g_{n}~z}{ z}{t_1}*t))
\end{array}
\]
Notice that by the base type interpretation, $f(g_1(x),g_2(x),...,g_n(x))=f_1'(g_1(x),g_2(x),...,g_n(x))*g_1'(x)+f_2'(g_1(x),g_2(x),...,g_n(x))*g_2'(x)+...+f_n'(g_1(x),g_2(x),...,g_n(x))*g_n'(x)$ where $f_j'$ means the derivative of $f$ to its $j$-th parameter, so we get the following and prove the case.
\[
\begin{array}{llll}
\myder{ N[(g_1~x)/y_1,(g_2~x)/y_2,...(g_{n}~x)/y_{n}]}{ x}{t_1}*t\\
\qquad=(\myder{ N[y_1'/y_1,g_2~t_1/y_2,...,g_{n}~t_1/y_{n}]}{ y_1'}{g_1~t_1}*(\myder{ g_1~z}{ z}{t_1}*t))\oplus ... \oplus \\
\qquad\qquad\qquad(\myder{  N[g_1~t_1/y_1,g_2~t_1/y_2,...,y_{n}'/y_{n}]}{ y_{n}'}{g_{n}~t_1}*(\myder{ g_{n}~z}{ z}{t_1}*t))
\end{array}
\]

\smallskip

\item Case $T_2$ is base type, $T_1$ is any type, $T$ is $A\rightarrow B$. We prove that for any $u$ of type $A$, we have $(\myder{ f(g~x)}{ x}{t_1}*t)~u=(\myder{ f~y}{ y}{(g~t_1)}*(\myder{ g~z}{ z}{t_1}*t))~u$.

First, let $f'=\lambda x:T_1.(f~x)~u$, $g'=g$, then by induction we have
\[
\begin{array}{llll}
\myder{ f'(g'~x)}{ x}{t_1}*t = \myder{ f'~y}{ y}{(g'~t_1)}*(\myder{ g'~z}{ z}{t_1}*t)
\end{array}
\]
that is, we have
\[
\begin{array}{llll}

(\myder{ f~(g~x)~u}{ x}{t_1}*t) = (\myder{ f~y~u}{ y}{(g~t_1)}*(\myder{ g~z}{ z}{t_1}*t))

\end{array}
\]

Then, we prove $(\myder{ f~(g~x)~u}{ x}{t_1}*t)=(\myder{ f~(g~x)}{ x}{t_1}*t)~u$ by the following calculation.
\[
\begin{array}{llll}
(\myder{ f~(g~x)}{ x}{t_1}*t)~u &=& (\myder{ \lambda a:A. (f~(g~x))~a}{ x}{t_1}*t)~u\\
 &=& (\lambda a:A. (\myder{ (f~(g~x))~a}{ x}{t_1}*t))~u\\
 &=& (\myder{ f~(g~x)~u}{ x}{t_1}*t)
\end{array}
\]

Next, we prove $(\myder{ f~y}{ y}{(g~t_1)}*(\myder{ g~z}{ z}{t_1}*t))~u=\myder{ f~y~u}{ y}{(g~t_1)}*(\myder{ g~z}{ z}{t_1}*t)$. For simplicity, we assume $T_1$ to be $(B,B,B,...,B)$ of $n$-tuples (the technique below can be applied to any $T_1$ type which makes the term well-typed).

On one hand, by substituting  $(g_1~z,g_2~z,...,g_{n}~z)$ for $g~z$, we have
\[
\begin{array}{llll}
(\myder{ f~y}{ y}{(g~t_1)}*(\myder{ g~z}{ z}{t_1}*t))~u\\
\qquad =  \myder{ f~y}{ y}{(g_1~t_1,g_2~t_1,...,g_{n}~t_1)}*(\myder{ (g_1~z,g_2~z,...,g_{n}~z)}{ z}{t_1}*t)~u\\
\qquad =(\myder{ f(y_1,g_2~t_1,...,g_{n}~t_1)}{ y_1'}{g_1~t_1}*(\myder{ g_1~z}{ z}{t_1}*t)\oplus ... \oplus \\
\qquad\qquad\qquad \myder{ f(g_1~t_1,g_2~t_1,...,y_{n}) }{ y_{n}'}{g_{n}~t_1}*(\myder{ g_{n}~z}{ z}{t_1}*t))~u
\end{array}
\]
Since
\[
\begin{array}{llll}
f(g_1~t_1,g_2~t_1,...,g_{j-1}~t_1,y_j,g_{j+1}~t_1,...,g_{n}~t_1)\\
\qquad= \lambda a:A.f~(g_1~t_1,g_2~t_1,...,g_{j-1}~t_1,y_j,g_{j+1}~t_1,...,g_{n}~t_1)~a
\end{array}
\]
which will be denoted as $ \lambda a:A.t_j^*$, we continue the calculation as follows.
\[
\begin{array}{llll}
(\myder{ f~y}{ y}{(g~t_1)}*(\myder{ g~z}{ z}{t_1}*t))~u\\
\qquad =(\myder{ \lambda a:A.t_1^*}{ y_1}{g_1~t_1}*(\myder{ g_1~z}{ z}{t_1}*t)\oplus ... \oplus \myder{ \lambda a:A.t_{n}^* }{ y_{n}}{g_{n}~t_1}*(\myder{ g_{n}~z}{ z}{t_1}*t))~u \\
\qquad = \lambda a:A. (\myder{ t_1^*}{ y_1}{g_1~t_1}*(\myder{ g_1~z}{ z}{t_1}*t))\oplus...\oplus (\myder{ t_{n}^*}{ y_n}{g_{n}~t_1}*(\myder{ g_{n}~z}{ z}{t_1}*t))~u\\
\qquad  = \myder{ t_1^*[u/a]}{ y_1}{g_1~t_1}*(\myder{ g_1~z}{ z}{t_1}*t)\oplus...\oplus \myder{ t_{n}^*[u/a]}{ y_n}{g_{n}~t_1}*(\myder{ g_{n}~z}{ z}{t_1}*t)
\end{array}
\]

On the other hand, we have
\[
\begin{array}{llll}
(\myder{ f~y~u}{ y}{(g~t_1)}*(\myder{ g~z}{ z}{t_1}*t))\\
\qquad =  \myder{ f~y~u}{ y}{(g_1~t_1,g_2~t_1,...,g_{n}~t_1)}*(\myder{ (g_1~z,g_2~z,...,g_{n}~z)}{ z}{t_1}*t)\\
\qquad =\myder{ f~(y_1,g_2~t_1,...,g_{n}~t_1)~u}{ y_1}{g_1~t_1}*(\myder{ g_1~z}{ z}{t_1}*t)\oplus ... \oplus \\
\qquad\qquad\qquad \myder{ f~(g_1~t_1,g_2~t_1,...,y_{n})~u }{ y_{n}}{g_{n}~t_1}*(\myder{ g_{n}~z}{ z}{t_1}*t)\\
\qquad =\myder{ (\lambda a:A.t_1^*)~u}{ y_1}{g_1~t_1}*(\myder{ g_1~z}{ z}{t_1}*t)\oplus ... \oplus \myder{ (\lambda a:A.t_{n}^*)~u }{ y_{n}}{g_{n}~t_1}*(\myder{ g_{n}~z}{ z}{t_1}*t)\\
\qquad = (\myder{ t_1^*[u/a]}{ y_1}{g_1~t_1}*(\myder{ g_1~z}{ z}{t_1}*t))\oplus...\oplus (\myder{ t_{n}^*[u/a]}{ y_n}{g_{n}~t_1}*(\myder{ g_{n}~z}{ z}{t_1}*t))
\end{array}
\]

Therefore, we \xmodify{prove}{have proven} the case.

\smallskip

\item Case $T_2$ is base type, $T_1$ is any type, $T$ is $(T_1,T_2,...,T_n)$. We need to prove that for all $j$, we have $\pi_j(\myder{ f(g~x)}{ x}{t_1}*t)=\pi_j(\myder{ f~y}{ y}{(g~t_1)}*(\myder{ g~z}{ z}{t_1}*t))$.
We may follow the proof for the case when T has type $A \rightarrow B$. Let $f' =\lambda x:T_1.\,\pi_j(f~x),g'=g$, by induction, we have
\[
\begin{array}{llll}

\myder{ \pi_j(f(g~x))}{ x}{t_1}*t = \myder{ \pi_j(f~y)}{ y}{(g~t_1)}*(\myder{ g~z}{ z}{t_1}*t)

\end{array}
\]
The rest of the proof is similar to that for the case when $T=A \rightarrow B$.

\item Case $T_2$ is base type, $T_1$ is any type, $T$ is $T_1+T_2$. Notice that $T_1$ has to be base type to be well-typed. But either the case, the proof is similar to the case when $T=A \rightarrow B$.

\smallskip

\item Case $T_2$ ,  $T_1$ and $T$ are any type. Notice that $T_2$ does not contain no $\rightarrow$ or $+$ to be well-typed (i.e., no derivative over function types). We have proved the case when $T_2$ is base type, and we assume that $T_2$ has type $(T_1,T_2,...,T_n)$.
Suppose the normal form of $t_1$ is $(t_{11}',t_{12}',...,t_{1n}')$ and the normal form of t is $(t_{21}',t_{22}',...,t_{2n}')$, Then
\[
\begin{array}{llll}
\myder{ f(g~x)}{ x}{t_1}*t\\
\qquad=\myder{ f(g~x)}{ x}{(t_{11}',t_{12}',...,t_{1n}')}*(t_{21}',t_{22}',...,t_{2n}')\\
\qquad=(\myder{ f(g~(x_1,t_{12}',...,t_{1n}'))}{ x_1}{t_{11}'},...,\myder{ f(g~(t_{11}',t_{12}',...,x_n))}{ x_{n}}{t_{1n}'})*(t_{21}',...,t_{2n}')\\
\qquad=(\myder{ f(g~(x_1,t_{12}',...,t_{1n}'))}{ x_1}{t_{11}'}*t_{21}')\oplus ...\oplus (\myder{ f(g~(t_{11}',t_{12}',...,x_n))}{ x_{n}}{t_{1n}'}*t_{2n}')
\end{array}
\]
On the other hand, we can use Lemma \ref{EqDist} (i.e., $t_1*(t_2\oplus t_3) = (t_1*t_2)\oplus (t_1*t_3)$) to do the following calculation.
\[
\begin{array}{llll}
\myder{ f~y}{ y}{(g~t_1)}*(\myder{ g~z}{ z}{t_1}*t)\\
\qquad=\myder{ f~y}{ y}{(g~t_1)}*((\myder{ g~(x_1,t_{12}',...,t_{1n}')}{ x_1}{t_{11}'}*t_{21}')\oplus ...\oplus (\myder{ g~(t_{11}',t_{12}',...,x_n)}{ x_{n}}{t_{1n}'}*t_{2n}'))\\
\qquad=\myder{ f~y}{ y}{(g~t_1)}*(\myder{ g~(x_1,t_{12}',...,t_{1n}')}{ x_1}{t_{11}'}*t_{21}')\oplus ...\oplus\\
\qquad\qquad\qquad \myder{ f~y}{ y}{(g~t_1)}*(\myder{ g~(t_{11}',t_{12}',...,x_n)}{ x_{n}}{t_{1n}'}*t_{2n}')
\end{array}
\]

Now by induction using $f'=f, g'=\lambda x:T_j.g~(t_{11}',t_{12}',...,t_{1(j-1)}',x,t_{1(j+1)}',...,t_{1n}')$, we have
\[
\begin{array}{llll}
\myder{ f(g~(t_{11}',t_{12}',...,t_{1(j-1)}',x_j,t_{1(j+1)}'...,t_{1n}')}{ x_j}{t_{1j}'}*t_{2j}'\\
\qquad = \myder{ f~y}{ y}{(g'~t_{1j}')}*(\myder{ g~(t_{11}',t_{12}',...,t_{1(j-1)}',x_j,t_{1(j+1)}'...,t_{1n}')}{ x_j}{t_{1j}'}*t_{2j}')\\
\qquad = \myder{ f~y}{ y}{(g~t_1)}*(\myder{ g~(t_{11}',t_{12}',...,t_{1(j-1)}',x_j,t_{1(j+1)}'...,t_{1n}')}{ x_j}{t_{1j}'}*t_{2j}')
\end{array}
\]
Therefore by Lemma \ref{EqAdd}, we \xmodify{prove}{have proven} the case.
\end{itemize}
\xmodify{none}{Thus we have proven the theorem.}
\end{proof}

\subsection*{Application: Automatic Differentiation}

The Chain Rule provides another way to compute the derivatives. There are many applications of the chain rule, and here we give an example of how to associate it with the \xmodify{auto}{automatic} differentiation \cite{Elli18}.

\begin{example}[AD] This is an example from \cite{Elli18}. Let $\m{sqr}$ and $\m{magSqr}$ be defined as follows.
\[
\begin{array}{llll}
\m{\m{sqr}} &::& \mathbb{R} \to \mathbb{R}\\
\m{\m{sqr}}~a &=& a * a\\
\m{\m{magSqr}} &::& (\mathbb{R},\mathbb{R}) \to \mathbb{R} \\
\m{\m{magSqr}}~(a, b) &=& \m{\m{sqr}}~a \oplus \m{\m{sqr}}~ b\\
\end{array}
\]
\end{example}
First of all, let $t_1$ and $t_2$ two pairs, then it is easy to prove that $\myder{ (t_1\oplus t_2)}{ x}{t_3}=\myder{ t_1}{ x}{t_3}\oplus \myder{ t_2}{ x}{t_3}$.
Next, we can perform automatic differentiation on $\m{magSqr}$ by the following calculation.
\[
\begin{array}{llll}
 \myder{ (\m{magSqr}~x)}{ x}{(a,b)}*t\\
\qquad = \myder{ (\m{sqr} (\pi_1 x) \oplus \m{sqr} (\pi_2 x))}{ x}{(a,b)}*t\\
\qquad = \myder{ (\m{sqr}~y)}{ y}{\pi_1 (a,b)}*(\myder{ (\pi_1 x)}{ x}{(a,b)}*t) \oplus \myder{ (\m{sqr}~y)}{ y}{\pi_2 (a,b)}*(\myder{(\pi_2 x)}{ x}{(a,b)}*t)\\
\qquad =2*a*((1,0)*t)\oplus 2*b*((0,1)*t)
\end{array}
\]

Now, because the theorem applies for any t of pair type, we use  $(1,0)$ and $(0,1)$ to substitute for $t$ respectively, and we will get $\myder{(\m{magSqr}~x)}{ x}{(a,b)}=(2*a,2*b)$, which means its derivative to $a$ is $2*a$ and its derivative to $b$ is $2*b$.

\section{Taylor's Theorem}
\label{sec:taylor}

In this section, we discuss Taylor's Theorem, which is useful to give an approximation of a $k$-order differentiable function around a given point by a polynomial of degree $k$.
In programming, it is important and has many applications in approximation and incremental computation. We first give an example and then we prove the theorem.

First of all, we introduce some high-order notations.
\[
\begin{array}{lrclll}
\frac{\partial^0 t_1}{\partial x^0}|_{t_2} &=& t_1&\quad \frac{\partial^n t_1}{\partial x^n}|_{t_2} &=&
\frac{\partial\frac{\partial^{n-1} t_1}{\partial x^{n-1}}|_{x}}{\partial x}|_{t_2}\\
t*t_1^0 &=& t&\quad t*t_1^n &=& (t*t_1)*t_1^{n-1}\\
f^{0}  &=& f&\quad f^{n} &=& (f')^{n-1}\\
(\lambda x:T.\,t)'&=& \lambda x:T.\,\frac{\partial t}{\partial x}|_{x}& & &
\end{array}
\]

\ignore{
\begin{definition}
 $(\lambda x:T.t)'=\lambda x:T. \frac{\partial t}{\partial x}|_{x}$ (notice that now t has no binding with the x of $\lambda x:T$, because the binding has changed)
\end{definition}

Notation: for $k\in$ base type, $kt = t * k$, $\frac{\partial^n t_1}{\partial x^n}|_{t_2}= \frac{\partial\frac{\partial^{n-1} t_1}{\partial x^{n-1}}|_{x}}{\partial x}|_{t_2}$, $\frac{\partial^0 t_1}{\partial x^0}|_{t_2}=t_1$, $t*t_1^n = (t*t_1)*t_1^{n-1}$, $t*t_1^0=t$, $f^{n}  = (f')^{n-1}$, $f^{0}  = f$
}

\begin{example}[Taylor]Consider a function $f$ on real numbers, usually defined in mathematics as $f (x,y) = (2*x*y, 3*x*x+y)$. In our calculus, it is defined as follows.

\[
\begin{array}{llll}

 \m{f} &::& (\mathbb{R},\mathbb{R})\to (\mathbb{R},\mathbb{R})\\
 \m{f}  &=& \lambda x:(\mathbb{R},\mathbb{R}).(2*\pi_1(x)*\pi_2(x),3*\pi_1(x)*\pi_1(x)\oplus \pi_2(x))\\

\end{array}
\]
The following expand the Taylor's theorem up to 2-order derivative.
\[
\begin{array}{llll}

 f~(C_1,C_2)&=& (2*C_1*C_2,3*C_1*C_1\oplus C_2)\\

 f~(0,0)&=& (0,0)\\

 f'~(0,0)*(C_1,C_2)&=&\reason{Application}\\   
&&\frac{\partial (2*\pi_1(x)*\pi_2(x),3*\pi_1(x)*\pi_1(x)\oplus \pi_2(x))}{\partial x}|_{(0,0)}*(C_1,C_2)\\

&=&\reason{Rule {\sc EAppDer3}}\\
&&(\frac{\partial (2*x_1*0,3*x_1*x_1\oplus 0)}{\partial x_1}|_{0},\frac{\partial (2*0*x_2,3*0*0\oplus x_2)}{\partial x_2}|_{0})*(C_1,C_2)\\

&=& \reason{Lemma \ref{EqLSub}}\\
&&((0,0),(0,1))*(C_1,C_2)\\

&=& \reason{Rule {\sc EAppMul4}, Rule {\sc EAppAdd1}}\\
&&(0,C_2)\\

\end{array}
\]
\[
\begin{array}{llll}

f''~(0,0)&=&\reason{Application}\\   
&&\frac{\partial\frac{\partial (2*\pi_1(x)*\pi_2(x),3*\pi_1(x)*\pi_1(x)\oplus \pi_2(x))}{\partial x}|_{x}}{\partial x}|_{(0,0)}\\

&=&\reason{Rule {\sc EAppDer3}}\\
&&(\frac{\partial\frac{\partial (2*\pi_1(x)*\pi_2(x),3*\pi_1(x)*\pi_1(x)\oplus \pi_2(x))}{\partial x}|_{(x_1,0)}}{\partial x_1}|_{0},\frac{\partial\frac{\partial (2*\pi_1(x)*\pi_2(x),3*\pi_1(x)*\pi_1(x)\oplus \pi_2(x))}{\partial x}|_{(0,x_2)}}{\partial x_2}|_{0})\\

&=&\reason{Rule {\sc EAppDer3}}\\
&& (\frac{\partial (\frac{\partial (2*x_1'*0,3*x_1'*x_1'\oplus 0)}{\partial x_1'}|_{x_1},\frac{\partial (2*x_1*x_2',3*x_1*x_1\oplus x_2')}{\partial x_2'}|_{0})}{\partial x_1}|_{0},\frac{\partial (\frac{\partial (2*x_1'*x_2,3*x_1'*x_1'\oplus x_2)}{\partial x_1'}|_{0},\frac{\partial (2*0*x_2',3*0*0\oplus x_2')}{\partial x_2'}|_{x_2})}{\partial x_2}|_{0})\\

&=&  \reason{Lemma \ref{EqLSub}}\\
&&(\frac{\partial ((0,6*x_1),(2*x_1,1))}{\partial x_1}|_{0},\frac{\partial ((2*x_2,0),(0,1))}{\partial x_2}|_{0})\\

&=& \reason{Lemma \ref{EqLSub}}\\
&& (((0,6),(2,0)),((2,0),(0,0)))\\

\end{array}
\]
\[
\begin{array}{llll}

(f''~(0,0))*(C_1,C_2)^2&=&\reason{Rule {\sc EAppMul4}, Rule {\sc EAppAdd1}}\\
&&((2*C_2,6*C_1),(2*C_1,0))*(C_1,C_2)\\

&=&\reason{Rule {\sc EAppMul4}}\\
&&(2*C_2*C_1,6*C_1*C_1)\oplus (2*C_1*C_2,0)\\

&=&\reason{Rule {\sc EAppAdd1}}\\
&&(4*C_1*C_2,6*C_1*C_1)\\

\end{array}
\]
\end{example}

Thus we have $f~(C_1,C_2)$ $=(2*C_1*C_2,3*C_1*C_1\oplus C_2)$ $=(0,0)\oplus(0,C_2)\oplus (2*C_1*C_2,3*C_1*C_1)$ $=f~(0,0)\oplus \frac{1}{1!}(f'~ (0,0))*(C_1,C_2) \oplus \frac{1}{2!}((f''~(0,0))*(C_1,C_2))*(C_1,C_2)$

\begin{theorem}[Taylor's Theorem]
  \label{theorem:taylor}
\rm
If both $f~ t$ and  $\sum\limits_{k=0}^{\infty}\frac{1}{k!}(f^{(k)}~t_0)*(t\ominus t_0)^k$ are weak-normalizable, then
\[
f~ t = \sum\limits_{k=0}^{\infty}\frac{1}{k!}(f^{(k)}~t_0)*(t\ominus t_0)^k.
\]
\end{theorem}

\begin{proof}

Like in the proof of Theorem \ref{theorem:nl}, for simplicity, we assume that $f$, $g$, $t$ and $t_1$ are closed. Furthermore, we assume that $t$ and $t_1$ are in normal form.
We prove it by induction on the type of $f : T \to T'$.

\begin{itemize}

\item Case $T'$ is a base type. $T$ must contain no $\rightarrow$ by our typing, so for simplicity, we suppose $T$ to be $(B,B,...,B)$.
%
Using the same technique in Theorem \ref{theorem:chainRule}, we assume $f$ to be
\[
f= \lambda x:T.\, (\lambda x_1:B.\,\lambda x_2:B.,...\lambda x_{n}:B.\,N)~\pi_1(x)~\pi_2(x)...~\pi_n(x)
\]
(denoted by $f= \lambda a:T.\, t_2$ later), $t$ to be $(t_{11},t_{12},...,t_{1n})$, and $t_0$ to be $(t_{21},t_{22},...,t_{2n})$, where each $t_{ij}$ is a normal form of base type. Then we have

\[
\begin{array}{llll}
(f^{(n)}~t_0)*(t\ominus t_0)^n\\\vspace{1ex}
\qquad =\frac{\partial^n t_2}{\partial x^n}|_{t_0}*(t\ominus t_0)^n\\\vspace{1ex}
\qquad =(\frac{\partial \frac{\partial^{n-1} t_2}{\partial x^{n-1}}|_{(x_1,t_{22},...,t_{2n})} }{\partial x_1}|_{t_{21}},...,\frac{\partial \frac{\partial^{n-1} t_2}{\partial x^{n-1}}|_{(t_{21},t_{22},...,x_{n})} }{\partial x_{n}}|_{t_{2n}})*(t\ominus t_0)^n\\\vspace{1ex}
\qquad =(\frac{\partial \frac{\partial^{n-1} t_2}{\partial x^{n-1}}|_{(x_1,t_{22},...,t_{2n})} }{\partial x_1}|_{t_{21}}*(t_{11}\ominus t_{21})\oplus...\oplus\frac{\partial \frac{\partial^{n-1} t_2}{\partial x^{n-1}}|_{(t_{21},t_{22},...,x_{n})} }{\partial x_{n}}|_{t_{2n}}*(t_{1n}\ominus t_{2n}))*(t\ominus t_0)^{n-1}\\\vspace{1ex}
\qquad =((\frac{\partial (\frac{\partial \frac{\partial^{n-2} t_2}{\partial x^{n-2}}|_{(x_1,t_{22},...,t_{2n})} }{\partial x_1}|_{x_1},...,\frac{\partial \frac{\partial^{n-2} t_2}{\partial x^{n-2}}|_{(x_1,t_{22},...,x_{n})} }{\partial x_{n}}|_{t_{2n}})}{\partial x_1}|_{t_{21}})*(t_{11}\ominus t_{21})\oplus...\oplus\\\vspace{1ex}
\end{array}
\]

\[
\begin{array}{llll}

\qquad \qquad \qquad (\frac{\partial (\frac{\partial \frac{\partial^{n-2} t_2}{\partial x^{n-2}}|_{(x_1,t_{22},...,x_n)} }{\partial x_1}|_{t_{21}},...,\frac{\partial \frac{\partial^{n-2} t_2}{\partial x^{n-2}}|_{(t_{21},t_{22},...,x_{n})} }{\partial x_{n}}|_{x_n})}{\partial x_n}|_{t_{2n}})*(t_{1n}\ominus t_{2n}))*(t\ominus t_0)^{n-1}\\\vspace{1ex}

\qquad =((\frac{\partial \frac{\partial \frac{\partial^{n-2} t_2}{\partial x^{n-2}}|_{(x_1,t_{22},...,t_{2n})} }{\partial x_1}|_{x_1}}{\partial x_1}|_{t_{21}})*(t_{11}\ominus t_{21})^2\oplus...\oplus\\\vspace{1ex}
\qquad \qquad \qquad  ((\frac{\partial \frac{\partial \frac{\partial^{n-2} t_2}{\partial x^{n-2}}|_{(x_1,t_{22},...,x_n)} }{\partial x_1}|_{t_{21}}}{\partial x_n}|_{t_{2n}})*(t_{1n}\ominus t_{2n}))*(t_{11}\ominus t_{21}),
\\\vspace{1ex}
\qquad\quad ((\frac{\partial \frac{\partial \frac{\partial^{n-2} t_2}{\partial x^{n-2}}|_{(x_1,x_2,...,t_{2n})} }{\partial x_2}|_{t_{22}}}{\partial x_1}|_{t_{21}})*(t_{11}\ominus t_{21}))*(t_{12}\ominus t_{22})\oplus...\oplus \\\vspace{1ex}
\qquad \qquad \qquad ((\frac{\partial \frac{\partial \frac{\partial^{n-2} t_2}{\partial x^{n-2}}|_{(t_{11},x_2,...,x_n)} }{\partial x_2}|_{t_{22}}}{\partial x_n}|_{t_{2n}})*(t_{1n}\ominus t_{2n}))*(t_{12}\ominus t_{22}),\\\vspace{1ex}
\qquad\quad...\\\vspace{1ex}
\qquad\quad ((\frac{\partial \frac{\partial \frac{\partial^{n-2} t_2}{\partial x^{n-2}}|_{(x_1,t_{22},...,x_n)} }{\partial x_n}|_{t_{2n}}}{\partial x_1}|_{t_{21}})*(t_{11}\ominus t_{21}))*(t_{1n}\ominus t_{2n})\oplus...\oplus \\\vspace{1ex}
\qquad \qquad \qquad (\frac{\partial \frac{\partial \frac{\partial^{n-2} t_2}{\partial x^{n-2}}|_{(t_{11},t_{22},...,x_n)} }{\partial x_n}|_{x_n}}{\partial x_n}|_{t_{2n}})*(t_{1n}\ominus t_{2n})^2)*(t\ominus t_0)^{n-2}\\\vspace{1ex}
\qquad =...
\end{array}
\]
As seen in the above, every time we decompose a $\frac{\partial }{\partial x_i}|_{(...)}$, apply Rule {\sc EAppDer1}, and then make reduction with Rule {\sc EAppMul3}
to lower down the exponent of $(t\ominus t_0)^n$.
Finally, we will decompose the last derivative and get the term $t_2$ in the form of
$t_2[t_{21}'/x_1,t_{22}'/x_2,...,t_{2n}'/x_{n}]$ where $\forall j\in[1,n],t_{2j}'$ is either $t_{2j}$ or $x_j$.

Note that on base type we assume that we have Taylor's Theorem:
\[
f(x_0+h)= f(x_0)+\sum\limits_{k=1}^{\infty}\frac{1}{k!}(\sum\limits_{i=1}^{n}h_i\frac{\partial}{\partial x_i})^k f(x_0)
\]
where $x_0$ and $h$ is an $n$-dimensional vector, and $x_j$, $h_j$ is its projection to its $j$-th dimension.

So we have $(f^{(k)}~t_0)*(t\ominus t_0)^k$ corresponds to the $k$-th addend $\frac{1}{k!}(\sum\limits_{i=1}^{n}h_i\frac{\partial}{\partial x_i})^k f(x_0)$.


\smallskip

\item Case: $T'$ is function type $A\rightarrow B$. Similar to the proof in Theorem \ref{theorem:chainRule}, for all $u$ of type $A$, we define $f^*=\lambda x:T.\,f~x~u$, and by using the inductive result on type $B$, we can prove the case simiarly as that  in Theorem \ref{theorem:chainRule}.

\smallskip

\item Case: $T'$ is a tuple type $(T_1,T_2,T_3,...)$. Just define $f^*=\lambda x:T.\,\pi_j(f~x)$ to use inductive result. The rest is simple.

\item Case: $T'$ is a tuple type $T_1+T_2$. This case is impossible because the righthand is not well-typed.
\end{itemize}
\xmodify{none}{Thus we have proven the theorem.}
\end{proof}
\subsection*{Application: Polynomial Approximation}

Taylor's Theorem has many applications. Here we give an example of using Taylor's Theorem for approximation. Suppose there is a point $(1,0)$ in the polar coordinate system, and we  want to know where the point will be if we slightly change the radius $r$ and the angle $\theta$. Since it is extremely costive to compute functions such as $\sin()$ and $\cos()$, Taylor's Theorem enables us to make a fast polynomial approximation.

\begin{example} Let function $\m{polar2catesian}$ be defined by
\[
\begin{array}{llll}
\m{polar2cartesian} &::& (\mathbb{R},\mathbb{R}) \to (\mathbb{R},\mathbb{R})\\
\m{polar2cartesian} (r,\theta) &=& (r* \cos(\theta),r*\sin(\theta))
\end{array}
\]
We show how to expand $\m{polar2cartesian}(r,\theta)$ at $(1,0)$ up to $2$nd-order derivative.
Since
\[
\begin{array}{llll}
\frac{\partial (\m{polar2cartesian}(x))}{\partial x}|_{(1,0)}
 &=\frac{\partial (\pi_1 x*\m{cos}(\pi_2 x),\pi_1 x*\m{sin}(\pi_2 x)}{\partial x}|_{(1,0)}\\
 &=(\frac{\partial (x_1*\cos(0),x_1*\sin(0))}{\partial x_1}|_{1},\frac{\partial (1*\cos(x_2),1*\sin(x_2))}{\partial x_2}|_{0})\\
 &=((1,0),(0,1))
\end{array}
\]
we have
\[
\begin{array}{llll}
\frac{\partial (\m{polar2cartesian}(x))}{\partial x}|_{(1,0)}*(\Delta r,\Delta \theta)
=(\Delta r,\Delta \theta).
\end{array}
\]
Again, we have
\[
\begin{array}{llll}
\frac{1}{2}\frac{\partial^2 (\m{polar2cartesian}(x))}{\partial x^2}|_{(1,0)}*(\Delta r,\Delta \theta)^2
 &=(((0,0),(0,1)),((0,1),(-1,0)))*(\Delta r,\Delta \theta)^2\\
 &=(-\frac{1}{2}\Delta \theta^2,\Delta r*\Delta \theta).
\end{array}
\]

Combining the above, we can use $(1\oplus\Delta r\ominus\frac{1}{2}\Delta \theta^2,\Delta \theta\oplus\Delta r*\Delta \theta)$ to make an approximation to $\m{\m{polar2cartesian}}(1+\Delta r,\Delta \theta)$.
\end{example}

\section{Discussion}
\label{sec:discusion}

In this section, we makes remarks on generality of our approach, and on how to deal with discrete derivatives in our context.

\subsection{More Theorems and Applications}


We keep many mathematical structures in our calculus. As a result, we can prove more theorems under this framework. We select the most important three, but there are many other theorems that hold in our system:
\begin{itemize}
\item$(t_1\ominus t_2)\oplus (t_2\ominus t_3)\oplus...\oplus(t_{n-1}\ominus t_n)=t_1\ominus t_n$,

\item$\myder{ t_1\oplus t_2}{ x}{t_3}=\myder{ t_1}{ x}{t_3}\oplus\myder{ t_2}{ x}{t_3}$,

\item$x\in B,\myder{ t_1*t_2}{ x}{t_3}=\myder{ t_1}{ x}{t_3}*t_2[t_3/x]\oplus t_1[t_3/x]*\myder{ t_2}{ x}{t_3}$,

\item$\forall t_1$, if $t_1$ contains no free $x$, $\myder{ t_1*x}{ x}{t_2}=t_1$.

\end{itemize}
Associated with each of these theorems is a bunch of applications. For lack of space, we only discuss three theorems in detail.

Now it is natural to ask whether all the theorems on base types have correspondence in our system. The answer is that it depends on the mathematical structure of the base types. In our proof, we assume the commutative law and associative law of addition and multiplication, and the distributive law of multiplication. We can construct a counterexample under this case.
Suppose there is a strange law on a base type that $\forall x,y, x*y=y$, which is interpreted by our system as $t_1*t_2=t_2$. Now let $t_1$  be $((r_1,r_2),(r_3,r_4))$, and $t_2$ be $(r_5,r_6)$. Then
\[
\begin{array}{llll}
t_1*t_2 &=& (r_1*r_5\oplus r_3* r_6,  r_2*r_5\oplus r_4* r_6)\\
        &=& (r_5\oplus r_6,r_5\oplus r_6)
\end{array}
\]
which does not equal to $t_2$. This means that our system does not preserve this strange law.

In our design of the calculus, we touch little on details of base types. So for some strange base types, we may not be able to preserve its mathematical structure. But as for the widely used $\mathbb{R}$ and $\mathbb{C}$, our system preserves most of their  important theorems.


It is interesting to note that it is impossible to prove these theorems using the theory of change \cite{CGRO14}, because the theory of change does not tell difference between smooth functions and non-continuous functions and use the same calculation for them. In our calculus, we distribute these calculation to base types step by step, and use these calculation (such as on base types, we have $\int_{a_1}^{a_2}\myder{ f~y}{ y}{x}=f~a_2\ominus f~a_1$) to prove our theorems.

\subsection{Discrete Derivatives}

\begin{figure}[t]
\[
\begin{array}{lrcll}
\mbox{Normal Form} & \nf &::=& \nb&\mynote{interpretable forms on base type}\\
 && | & \lambda x:T. t&\mynote{function, t can't be further reduced}\\
\\
\mbox{Interpretable Normal Forms}& \nb &::=& c & \mynote{constants of interpretable type}\\
 && | & \nb~\nf &\mynote{primitive functions application} \\
 && | & \nb \oplus  \nf \ ~~|~ \nf \oplus  \nb&\mynote{addition on interpretable type}\\
 && | & \nb \ominus \nf \ ~~|~ \nf \ominus \nb&\mynote{subtraction on interpretable type} \\
 && | & x      &\mynote{interpretable types variables} \\
 && | & \frac{\Delta \nb}{\Delta x}|_{\nf,\nf} &\mynote{derivative on base type}\\

\end{array}
\]
\caption{Discrete normal form}
\label{fig:term}
\end{figure}

We can define discrete version of our calculus, where we represent changes as discrete deltas instead of through derivatives and integrations. We will show the equivalence between our discrete version and change theory  \cite{CGRO14} by implementing function $Derive$ in our calculus.

The normal form this time is defined in Figure \ref{fig:term}. We use the term $\frac{\Delta t}{\Delta x}|_{t,t}$ to represent discrete derivative. This time we can easily manipulate values of base types because we only require the operator $\oplus$ and $\ominus$ to be well-defined. Also notice that this time we can implement derivatives on function type.

\ignore{
As for reduction rules, similar to

\reductiond

\noindent
we can define

\begin{prooftree}
\AxiomC{$t_0, t_1:\iB$}
\UnaryInfC{$\frac{\Delta \lambda y:T.t}{\Delta x}|_{t_0,t_1}\ \rightarrow \ \lambda y:T.(\frac{\Delta t}{\Delta x}|_{t_0,t_1})
$}
\end{prooftree}

To show the equivalence between the two systems, one direction is from change theory to our discrete version. Of course we can use the change theory to implement our system if the operator $\oplus$ is well-defined.
}

To show that our discrete version can be used to implement the change theory \cite{CGRO14}
it is sufficient to consider terms of base types or function types, without need to to consider tuples and the operator * and $\int$. We want to use our calculus to implement function $Derive$ which satisfies the equation $(Derive~f) x~\Delta x= f(x\oplus \Delta x)\ominus f(x)$.

For interpretation of derivatives on base types, we just require they satisfy  $\frac{\Delta t}{\Delta y}|_{t_1,t_2}$ = $t[t_1\oplus t_2/y]\ominus t[t_1/y]$. Then similarly to Newton-Leibniz Theorem we can prove $\frac{\Delta (f\,y)}{\Delta y}|_{x,\Delta x}$ = $f~ (x\oplus\Delta x)\ominus f~ x$ (where $f$ does not contain free $y$), which is our version of function $Derive$.

To see this clear, in change theory, we write function $Derive$ and the system will automatically calculate it by using the rules:
\[
\begin{array}{llll}
\m{Derive}~c &=& 0\\
\m{Derive}~x &=& \Delta x\\
\m{Derive}(\lambda x:T.t) &=&\lambda x:T.\,\lambda dx:\Delta T.\,\m{Derive}(t)\\
\m{Derive}(s~t) &=& \m{Derive}(s)~t~\m{Derive}(t)
\end{array}
\]
In our calculus, one writes $\frac{\Delta f~y}{\Delta y}|_{x,\Delta x}$, and the system will automatically calculate the following rules:
\[
\begin{array}{llll}
\frac{\Delta c}{\Delta y}|_{x,\Delta x} &=& 0\\
\frac{\Delta y}{\Delta y}|_{x,\Delta x} &=& \Delta x\\
\frac{\Delta \lambda y:T.t}{\Delta x}|_{t_0,t_1} &=& \lambda y:T.(\frac{\Delta t}{\Delta x}|_{t_0,t_1})\\
(\lambda x.\lambda \Delta x.\, \frac{\Delta t}{\Delta y}|_{x, \Delta x})~t_1~t_2
&=& \lambda y:T.\,t~(t_1\oplus t_2)\ominus\lambda y:T.\,t~t_1\\
\end{array}
\]
\ignore{
\begin{itemize}
\item Case $\m{Derive}~c=0$, we have $\frac{\Delta c}{\Delta y}|_{x,\Delta x}$ = 0.

\item Case $\m{Derive}~\Delta x$, we have $\frac{\Delta y}{\Delta y}|_{x,\Delta x}$ = $\Delta x$.

\item Case $\m{Derive}(\lambda x:T.t)=\lambda x:T.\lambda dx:T.\m{Derive}(t)$, we have

\[
\begin{array}{llll}

\lambda x.\lambda \Delta x. \frac{\Delta t}{\Delta y}|_{x, \Delta x}~t_1~t_2\\

\qquad=\frac{\Delta t}{\Delta y}|_{t_1, t_2}\\

\qquad=t[t_1\oplus t_2/y]\ominus t[t_1/y]\\

\qquad=\lambda y:T.t~(t_1\oplus t_2)\ominus\lambda y:T.t~t_1

\end{array}
\]

\end{itemize}
}
Notice that the first three rules have good correspondence, while the last one is a bit different. This is because in change theory's definition, we have $\Delta (A\rightarrow B)=A\rightarrow \Delta A\rightarrow \Delta B$, while in our calulus, we have $\Delta (A\rightarrow B)= A\rightarrow \Delta B$. We, fortunately, can achieve the same effect through  Newton-Leibniz Formula.

\section{Related Work}
\label{sec:relatedWork}

\noindent
{\bf Differential Calculus and The Change Theory}~
The differential \xmodify{$\lambda$-calculus}{lambda-calculus} \cite{EhRe03,Ehrh18} has been studied for computing  derivatives of arbitrary higher-order programs. In the differential \xmodify{$\lambda$-calculus}{lambda-calculus}, derivatives are guaranteed to be linear in its argument, where the incremental \xmodify{$\lambda$-calculus}{lambda-calculus} does not have this restriction. Instead, it requires that the function should be differentiable. The big difference between our calculus and differential lambda calculus is that we perform computation on terms instead of analysis on terms.

The idea of performing incremental computation using derivatives has been studied by Cai et al. \cite{CGRO14}, who give an account using change structures. They use this to provide a framework for incrementally evaluating lambda calculus programs. It is shown that the work can be enriched with recursion and fix-point computation \cite{Alva19}. The main difference between our work and \xmodify{the change theory}{change theory} is that we describe changes as mathematical derivatives while the change theory describe changes as (discrete) deltas.

\medskip

\noindent
{\bf Incremental/Self-Adaptive Computation}~
Paige and Koenig \cite{PaKo82} present derivatives for a first-order language with a fixed set of primitives for incremental computation. Blakeley et al. \cite{LaZh07} apply these ideas to a class of relational queries. Koch \cite{Koch10} guarantees asymptotic speedups with a compositional query transformation and delivers huge speedups in realistic benchmarks, though still for a first-order database language. We have proved \xmodify{the Taylor's theorem}{Taylor's theorem} in our framework, which provides us with another way to perform finite difference on the computation.

Self-adjusting computation \cite{AcAB08} or adaptive function programming \cite{AcBH02} provides a dynamic approach to incrementalization. In this approach, programs execute on the original input in an enhanced runtime environment that tracks the dependencies between values in a dynamic dependence graph; intermediate results are memoized. Later, changes to the input propagate through dependency graphs from changed inputs to results, updating both intermediate and final results; this processing is often more efficient than recomputation. Mathematically, \xmodify{the self-adjusting computation}{self-adjusting computations} corresponds to differential \xmodify{equation}{equations} (The \xmodify{change rate (or derivative)}{derivative} of a function can be represented by the computational result of function), which may be a future work of our calculus.

\medskip

\noindent
{\bf Automatic Differentiation}~
Automatic differentiation \cite{GrWa08} is a technique that allows for efficiently computing the derivative of arbitrary programs, and can be applied to probabilistic modeling \cite{Kucu17} and machine learning \cite{Bayd17}.
This technique has been successfully applied to some higher-order languages \cite{SiPe08,Elli18}. As pointed out in \cite{Alva19},
while some approaches have been suggested \cite{Manz12,KePS16}, a general theoretical framework for this technique is still a matter of open research. We prove the chain rule inside our framework, which lays a foundation for our calculus to perform automatic differentiation. And with more theorems in our calculus, we expect more profound \xmodify{application}{applications} in differential calculus.

\section{Conclusion}
\label{sec:conclusion}

In this paper, we propose an analytical differential calculus which is equipped with integration. This calculus, as far as we are aware, is the first one that has well-defined integration, which has not appeared in both differential lambda calculus and the change theory. Our calculus enjoys many nice properties such as soundness and strong normalizing (when $fix$ is excluded), and has three important theorems, which have profound applications in computer science. We believe the following directions will be important in our future work.

\begin{itemize}
\item {\em Adding more theorems.} We may wish to write programs on many specialized base types besides $\mathbb{R}$ and $\mathbb{C}$. As we have demonstrated in this paper, our calculus preserves many important computational structures on base types. Therefore, it is possible to extend our system with theorems having ome unique mathematical structures and use these theorems to optimize computation.

\item {\em Working on Derivatives on functions.}
We did not talk about derivatives on continuous functions because we have not had a good mathematical definition for them from perspective of computation. But derivatives on functions would be useful; it would be nice if we could use $\int_{\oplus}^{*}\myder{x(a_1,a_2)}{x}{x}dx$ to compute $a_1*a_2\ominus (a_1\oplus a_2)$.

\item {\em Manipulating differential equations}.
Differential equations would be very useful for users to program dynamic systems directly; one may write differential equations on data structures without writing the primitive forms of functions. It could be applied in many fields such as self-adjusting computation or self-adaptive system construction.

\end{itemize}



\bibliography{reference}

\newpage
\appendix
\section{Appendix-Calculus Property}
\subsection{Progress}

\begin{lemma}[Progress]\label{ProgressProof}\rm
   Suppose t is a well-typed term (Allow free variables of interpretable type $\iB$), then t is either a normal form or there is some $t'$ \xmodify{that}{such that} $t$$\rightarrow$$t'$.
\end{lemma}

\begin{proof}

We prove this by induction on form of $t$.
\begin{itemize}
\item Case $c$. \\
It is a normal form.

\smallskip

\item Case $t_1\oplus t_2$. \\
$t$ is well-typed if and only if $t_1$ has the same type T with $t_2$.
If $t_1$ or $t_2$ is not a normal form, we make reductions on  $t_1$ or $t_2$.
\xmodify{If both $t_1$ and $t_2$ are normal forms, if}{If} either $t_1$ or $t_2$ is \textbf{nb}, then $t_1 \oplus t_2$ is a \textbf{nb}\xmodify{, for other}{. For other} cases of normal forms, we have
\[
\begin{array}{rcl}
  (t_{11},t_{12},...t_{1n}) \oplus (t_{21},t_{22},...t_{2n})  &\rightarrow& (t_{11}\oplus t_{21},t_{12}\oplus t_{22},...t_{1n}\oplus t_{2n})\\
(\lambda x: T.t_1) \oplus (\lambda y: T.t_2)
  &\rightarrow& \lambda x: T.t_1 \oplus  (t_2[x/y])
\end{array}
\]

\smallskip

\item Case $t_1 \ominus t_2$.\\
 It is the same case with the $t_1 \oplus t_2$.

\smallskip

\item Case $x$. \\
Then $x$ is an interpretable type free variable, otherwise it is not well-typed. \xmodify{and an}{An} interpretable type free variable is a normal form.

\smallskip

\item Case $inl/inr~t$. \\
\xmodify{if}{If} $t$ is not a normal form, then we can make reduction in $t$, else this term itself is a normal form.

\smallskip

\item Case $case~t~of~inl~x_1 \Rightarrow t_1|~inr~x_2 \Rightarrow t_2$. \\
To be well-typed t has to be the type of $T_1+T_2$\xmodify{, if}{. If} $t$ is not a normal form, then we can make reduction in $t$, else $t$ has to be $inl/inr~t'$. So we can make reduction to $t_1[t'/x_1]$ or $t_2[t'/x_2]$

\smallskip

\item Case $\lambda x:T. t$. \\
It is a normal form if t can't be further reduced.

\smallskip

\item Case $t_1~t_2$.\\
If $t_1$ is not a normal form then we make reductions on $t_1$.

If $t_1$ is a normal form, then $t_1$ has to be $\lambda x:T. t$, or \textbf{nb}. For the former case we have
\[
(\lambda x:T. t)t_1  \rightarrow t[t_1/x]
\]

For the latter case, $t_2$ must be a \textbf{nf}, or it can make further reductions. So $t_1~t_2$ is a \textbf{nb}.

\smallskip

\item Case $\int_{t_1}^{t_2} t_3dx$.\\
If $t_1$ or $t_2$ is not a normal form then we can make reductions on $t_1$ or $t_2$.

If both $t_1$ and $t_2$ are normal forms, then \xmodify{$t_1$,$t_2$}{$t_1$ and $t_2$} have to be $(\textbf{nf},\textbf{nf},..\textbf{nf})$ or base type to be well-typed. \xmodify{if}{If} it is the former case.
\[
\begin{array}{llll}
\int_{(t_{11},t_{12},...t_{1n}) }^{(t_{21},t_{22},...,t_{2n})}tdx~
  & \rightarrow & \int_{t_{11}}^{t_{21}}  \pi_1(t[(x_1,t_{12},...t_{1n})/x])dx_1 ~\oplus \\
  & &  \int_{t_{12}}^{t_{22}}  \pi_2(t[(t_{21},x_2,...,t_{1n})/x])dx_2 ~\oplus \\
  & & \qquad \vdots  \\
  & & \int_{t_{1n}}^{t_{2n}} \pi_n(t[(t_{21},t_{22},...,x_{n})/x])dx_{n}
\end{array}
\]

If it is the latter case, let us inspect $t_3$. \xmodify{if}{If} $t_3$ is not a normal form, then we can make reductions on $t_3$ (notice that we only introduce a base type free variable into $t_3$).

If $t_3$ is a normal form, \xmodify{if}{and if} $t_1$,$t_2$ and $t_3$ are \textbf{nb}, then $\int_{t_1}^{t_2}t_3dx~$ is a normal form\xmodify{, for}{. For} other cases of normal forms:

\reductione

\reductionf

\reductions
\smallskip

\item Case $(t_1,t_2,...,t_{n})$.\\
If $t_i$ is not a normal form, then we make reductions on $t_i$.
If all the $t_i$ are normal forms, then t is a normal form.

\smallskip

\item Case $\pi_j(t_1)$.\\
If $t_1$ is not a normal form, then we make reductions on $t_1$.

If $t_1$ is a normal form, then it has to be $(\textbf{nf},\textbf{nf},...,\textbf{nf})$ to be well-typed, then we have
\[
\pi_j(t_1',t_2',...t_{n}')  \rightarrow t_j'
\]

\smallskip

\item Case $\myder{t_1}{x}{t_2}$.\\
If $t_2$ is not a normal form, then we make reductions on $t_2$.

If $t_2$ is a normal form, then it has to be $(t_1,...,t_{n})$ or an \textbf{nb}. \xmodify{if}{If} it is the form case:

\reductiong

If it is the latter case, if $t_1$ is not a normal form, then we can make reductions on $t_1$ (notice that we only introduce a base type free variable into $t_1$)\xmodify{, if}{. If} $t_1$ is a \textbf{nb}, then t is a \textbf{nb}, else we have

\reductionc

\reductiond

\reductionr
\smallskip

\item Case $t_1*t_2$.\\
If $t_1$ or $t_2$ is not a normal form, then we can make reductions on $t_1$ or $t_2$.

If  both $t_1$ and $t_2$ are normal forms, $t_2$ has to be $(t_1,...,t_{n})$ or a \textbf{nb}. \xmodify{if}{If} it is the former case, $t_1$ has also to be $(t_1,...,t_{n})$, then we have

\reductiono

If $t_2$ is a \textbf{nb}, if $t_1$ is a \textbf{nb}, then $t_1*t_2$ is a \textbf{nb}, else we have

\reductionm
\reductionn
\reductionv

\item Case $fix~f$.\\
Then we have $fix~f \rightarrow f~(fix~f)$

\end{itemize}
\end{proof}

\subsection{Preservation}

\begin{lemma}[Preservation under substitution]\label{Pns}\rm
  If $\Gamma,x:S\vdash t:T$ and $\Gamma\vdash s:S$, then we have $\Gamma\vdash t[s/x]:T$.
\end{lemma}

\begin{proof}

First we prove preservation under substitution.

\begin{itemize}
\item Case c.\\
Then $c[s/x]$ is $c$, therefore $\Gamma\vdash t[s/x]:$\basetype
\xmodify{fix the font}{}
\item Case $t_1 \oplus t_2$.\\
\ \ Suppose $\Gamma,x:S\vdash t_1 \oplus t_2:T$, then we have $\Gamma,x:S\vdash t_1:T,t_2:T$, based on induction we have $\Gamma\vdash t_1[s/x] \oplus t_2[s/x]:T$, therefore $\Gamma\vdash (t_1 \oplus t_2)[s/x]:T$.

\ \ Using the same techniques we can prove the case of $t_1 \ominus t_2$, $t_1 *t_2$, $t_1~t_2$, $\lambda x:T.t$, $\myder{t_1}{x}{t_2}$, $\int_{t_1}^{t_2}t_3dx$, $(t_1,t_2,...,t_n)$, $\pi_j(t)$ and $fix~f$, $inl/inr~t$, $case~t~of~inl~x_1 \Rightarrow t_1|~inr~x_2 \Rightarrow t_2$.

\smallskip

\item Case y.\\
\ \ \xmodify{if}{If} y = x then y[s/x] = s, so $\Gamma\vdash y[s/x]:T$.

\ \ \xmodify{if}{If} y is other than x, then y[s/x] = y, so $\Gamma\vdash y[s/x]:T$.
\end{itemize}
Then we prove the preservation

\xmodify{moved here}
{\begin{lemma}[Preservation]\label{PreservationProof}\rm
If $t : T$ and $t \rightarrow t'$, then $t' : T$. (Allowing free variable of $\iB$)
\end{lemma}
}

\begin{itemize}

\item Case $(\lambda x:T. t)t_1  \rightarrow t[t_1/x]$: It is straightforward by using the \xmodify{Lemma. (Preservation under substitution)}{Lemma \ref{Pns}.}\\

\smallskip

\item Case $fix~f\rightarrow f~(fix~f)$ \\
\ \ \ \ Suppose $\Gamma\vdash f:A\rightarrow A$, then $\Gamma\vdash fix~f:A$ and $\Gamma\vdash f~(fix~f):A$, so they have the same type.

\smallskip

\item Case $\pi_j(t_1,t_2,...t_{n})  \rightarrow t_j$ \\
\ \ \ \ Suppose $\Gamma\vdash (t_1,t_2,...t_{n}):(T_1,T_2,...,T_n)$, then $\Gamma\vdash \pi_j(t_1,t_2,...t_{n}):T_j$ and $\Gamma\vdash t_j:T_j$, so they have the same type.

\smallskip

\item Case
\reductionc

Suppose $\Gamma\vdash \myder{(t_1, t_2,...,t_{n})}{x}{t_0}:(T_1,T_2,..,T_{n})$, Then, $\Gamma,x:$\basetype$\vdash t_j:T_j$, then $\Gamma\vdash (\myder{t_1}{x}{t_0},\myder{t_2}{x}{t_0},...,\myder{t_{n}}{x}{t_0}):(T_1,T_2,..,T_{n})$, so they have the same type.

Using the same technique, we can prove the case

\reductionn

\reductione

\smallskip

\item Case \reductiond

Suppose $\Gamma\vdash \myder{\lambda y:T. t}{x}{t_0}:A \rightarrow B$, then $\Gamma,y:A\vdash\myder{t}{x}{t_0}:B$, therefore $\Gamma\vdash \lambda y:T .\myder{t}{x}{t_0}:A\rightarrow B$, so they have the same type.

Using the same techniques, we can prove the case

\reductionf

\reductionm

\smallskip

\item Case $(t_{11},t_{12},...t_{1n}) \oplus (t_{21},t_{22},...t_{2n})  \rightarrow (t_{11}\oplus t_{21},t_{12}\oplus t_{22},...t_{1n}\oplus t_{2n})$ \\

From $\Gamma \vdash (t_{11},t_{12},...t_{1n}) \oplus (t_{21},t_{22},...t_{2n}):T $, we have $\Gamma \vdash (t_{11},t_{12},...t_{1n}):T, (t_{21},t_{22},...t_{2n}):T$. Suppose T is $(T_1,T_2,...,T_{n})$, then we have $\Gamma \vdash t_{1i}:T_i,~t_{2i}:T_i$. \xmodify{therefore}{Therefore} $\Gamma \vdash t_{1i}\oplus t_{2i}:T_i$. So we have $\Gamma \vdash (t_{11}\oplus t_{21},t_{12}\oplus t_{22},...t_{1n}\oplus t_{2n}) : (T_1,T_2,...,T_{n}) = T$.

\ \ \ \

Using the same techniques, we can prove the preservation of the following rules

\ \ \ \ \ \ \ \ \ \ \ \ \ \ \ \ \ \ \ \ \ \ \ \ \ \ \ \ \ \ \ \ \ \ \ \ \reductionj

\ \ \ \ \ \ \ \ \ \ \ \ \ \ \ \ \ \ \ \ \ \ \ \ \ \ \ \ \ \ \ \ \ \ \ \ \reductionk

\ \ \ \ \ \ \ \ \ \ \ \ \ \ \ \ \ \ \ \ \ \ \ \ \ \ \ \ \ \ \ \ \ \ \ \ \reductionl

and reduction for $case~t~of~inl~x_1 \Rightarrow t_1|~inr~x_2 \Rightarrow t_2$.

\smallskip

\item Case
\reductiong

Let's suppose $\myder{t}{x}{(t_1, t_2,...,t_{n})}$ has the type $\frac{\partial T}{\partial T_0}$, then we have $\Gamma,x:T_0\vdash t:T$. Suppose that $T_0 = (T_1,T_2,...,T_{n})$, then we \xmodify{got}{have} $\Gamma,x_i:T_i\vdash  t[t_{i*}/x]:T$. Thus $\Gamma\vdash \myder{t[t_{i*}/x]}{x_i}{t_i}:\frac{\partial T}{\partial T_i}$. Therefore $ (\myder{t[t_{1*}/x]}{x_1}{t_1},\myder{t[t_{2*}/x]}{x_2}{t_2},...,\myder{t[t_{n*}/x]}{x_{n}}{t_{n}}) : (\frac{\partial T}{\partial T_1},\frac{\partial T}{\partial T_2},...,\frac{\partial T}{\partial T_{n}})=\frac{\partial T}{\partial T_0}$.

\smallskip

\item Case\reductionh

 Let's suppose $\Gamma,x:T_0\vdash t:\frac{\partial T}{\partial T_0}$ and $\Gamma\vdash\int_{(t_{11},t_{12},...t_{1n}) }^{(t_{21},t_{22},...,t_{2n})}tdx:T$.
Assume that $T_0 = (T_1,T_2,...,T_{n})$. Then for all j, we have $\Gamma,x_j:T_j\vdash t[(t_{21},...,t_{2(j-1)},x_j,t_{1(j+1)},...,t_{1n})/x ]:\frac{\partial T}{\partial T_0}$. Thus $\Gamma,x_j:T_j\vdash\int_{t_{1j}}^{t_{2j}}  \pi_j(t[(t_{21},...,t_{2(j-1)},x_j,t_{1(j+1)},...,t_{1n}) /x])dx_j :T_j $. Therefore $\Gamma\vdash \int_{t_{11}}^{t_{21}}  \pi_1(t[(x_1,t_{12},...t_{1n})/x_1])dx_1 \oplus  \int_{t_{12}}^{t_{22}}  \pi_2(t[(t_{21},x_2,...,t_{1n})/x_2])dx_2 \oplus ... \oplus\\ \int_{t_{1n}}^{t_{2n}} \pi_n(t[(t_{21},t_{22},...,x_{n})/x_{n}])dx_{n} :T$.

Using the same technique, we can prove the case.

\reductiono

Therefore, we prove the preservation of the system.
\end{itemize}
\end{proof}

\subsection{Confluence}

Define a binary relation $\twoheadrightarrow$ by induction on relation on terms.

$ M\twoheadrightarrow M$

\begin{prooftree}
\AxiomC{$  M\twoheadrightarrow M',N\twoheadrightarrow N' $}
\UnaryInfC{$M~N\twoheadrightarrow M'~N',M\oplus N\twoheadrightarrow M'\oplus N',M\ominus N\twoheadrightarrow M'\ominus N',M* N\twoheadrightarrow M'* N',\myder{M}{x}{N}\twoheadrightarrow \myder{M'}{x}{N'}$}
\end{prooftree}

\begin{prooftree}
\AxiomC{$   \forall j\in[1,n], M_j\twoheadrightarrow M_j'  $}
\UnaryInfC{$(M_1,M_2,..,M_{n})\twoheadrightarrow (M_1',M_2',..,M_{n}')$}
\end{prooftree}

\begin{prooftree}
\AxiomC{$    M_1\twoheadrightarrow M_1',M_2\twoheadrightarrow M_2',M_3\twoheadrightarrow M_3'  $}
\UnaryInfC{$\int_{M_1}^{M_2}M_3dx \twoheadrightarrow \int_{M_1'}^{M_2'}M_3'dx,case~M_1~of~inl~x_1 \Rightarrow M_2|~inr~x_2 \Rightarrow M_3 \twoheadrightarrow case~M_1'~of~inl~x_1 \Rightarrow M_2'|~inr~x_2 \Rightarrow M_3'$}
\end{prooftree}

\begin{prooftree}
\AxiomC{$    M_1\twoheadrightarrow M_1',M_2\twoheadrightarrow M_2',M_3\twoheadrightarrow M_3'  $}
\UnaryInfC{$case~inr~M_1~of~inl~x_1 \Rightarrow M_2|~inr~x_2 \Rightarrow M_3 \twoheadrightarrow M_3'[M_1'/x_2]$}
\end{prooftree}

\begin{prooftree}
\AxiomC{$    M_1\twoheadrightarrow M_1',M_2\twoheadrightarrow M_2',M_3\twoheadrightarrow M_3'  $}
\UnaryInfC{$case~inl~M_1~of~inl~x_1 \Rightarrow M_2|~inr~x_2 \Rightarrow M_3 \twoheadrightarrow M_2'[M_1'/x_1]$}
\end{prooftree}

\begin{prooftree}
\AxiomC{$    M\twoheadrightarrow M' $}
\UnaryInfC{$\lambda x:T.M\twoheadrightarrow \lambda x:T.M',\pi_j(M)\twoheadrightarrow \pi_j(M'),inl/inr~M\twoheadrightarrow inl/inr~M$}
\end{prooftree}

\begin{prooftree}
\AxiomC{$   \forall j\in[1,n], M_j\twoheadrightarrow M_j'  $}
\UnaryInfC{$\pi_j(M_1,M_2,..,M_{n})\twoheadrightarrow M_j'$}
\end{prooftree}

\begin{prooftree}
\AxiomC{$   M\twoheadrightarrow M',N\twoheadrightarrow N'  $}
\UnaryInfC{$(\lambda x:T.M)N \twoheadrightarrow M'[N'/x]$}
\end{prooftree}

\begin{prooftree}
\AxiomC{$ \forall j\in[1,n], M_{1j}\twoheadrightarrow M_{1j}',M_{2j}\twoheadrightarrow M_{2j}'$}
\UnaryInfC{$(M_{11},M_{12},...M_{1n}) \oplus (M_{21},M_{22},...M_{2n})  \twoheadrightarrow (M_{11}'\oplus M_{21}',M_{12}'\oplus M_{22}',...M_{1n}'\oplus M_{2n}')$}
\end{prooftree}

\begin{prooftree}
\AxiomC{$  \forall j\in[1,n], M_{1j}\twoheadrightarrow M_{1j}',M_{2j}\twoheadrightarrow M_{2j}'$}
\UnaryInfC{$(M_{11},M_{12},...M_{1n}) \ominus (M_{21},M_{22},...M_{2n})  \twoheadrightarrow (M_{11}'\ominus M_{21}',M_{12}'\ominus M_{22}',...M_{1n}'\ominus M_{2n}')$}
\end{prooftree}

\begin{prooftree}
\AxiomC{$   M\twoheadrightarrow M',N\twoheadrightarrow N'  $}
\UnaryInfC{$(\lambda x:T. M )\oplus (\lambda y:T. N) \twoheadrightarrow \lambda x:T. M' \oplus N'[y/x]$}
\end{prooftree}

\begin{prooftree}
\AxiomC{$   M\twoheadrightarrow M',N\twoheadrightarrow N'  $}
\UnaryInfC{$(\lambda x:T. M )\ominus (\lambda y:T. N) \twoheadrightarrow \lambda x:T. M' \ominus N'[y/x]$}
\end{prooftree}

\begin{prooftree}
\AxiomC{$ \forall j\in[1,n], M_j\twoheadrightarrow M_j',N\twoheadrightarrow N',N:\basetype$}
\UnaryInfC{$\myder{(M_1,M_2,..,M_{n})}{x}{N}\twoheadrightarrow (\myder{M_1'}{x}{N'},\myder{M_2'}{x}{N'},...,\myder{M_{n}'}{x}{N'})$}
\end{prooftree}

\begin{prooftree}
\AxiomC{$M\twoheadrightarrow M',N\twoheadrightarrow N',N:\basetype$}
\UnaryInfC{$\myder{\lambda y:T. M}{x}{N}\twoheadrightarrow\lambda y:T.\myder{M'}{x}{N'}$}
\end{prooftree}

\begin{prooftree}
\AxiomC{$  \forall j\in[1,n],  M_j\twoheadrightarrow M_j' ,(M_1', M_2'...,M_{j-1}',x_j,M_{j+1}'...,M_{n}')is~written~as~ M_{j*}',M_0\twoheadrightarrow M_0'$}
\UnaryInfC{$\myder{M_0}{x}{(M_1, M_2,...,M_{n})} \twoheadrightarrow (\myder{M_0'[M_{1*}'/x]}{x_1}{M_1'},\myder{M_0[M_{2*}'/x]}{x_2}{M_2'},...,\myder{t[M_{n*}'/x]}{x_{n}}{M_{n}'})$}
\end{prooftree}

\begin{prooftree}
\AxiomC{$M_0\twoheadrightarrow M_0',\forall j\in[1,n], M_{1j}\twoheadrightarrow M_{1j}',M_{2j}\twoheadrightarrow M_{2j}',(M_{11}'...,M_{1j-1}',x_j,M_{2j+1}'...,M_{2n}')is~written~as~ M_{j*}'$}
\UnaryInfC{$\int_{(M_{11},M_{12},...M_{1n}) }^{(M_{21},M_{22},...,M_{2n})}M_0dx~ \twoheadrightarrow \int_{M_{11}'}^{M_{21}'}  \pi_1(M_0'[M_{1*}'/x])dx_1 \oplus ... \oplus \int_{M_{1n}'}^{M_{2n}'} \pi_n(M_0'[M_{n*}'/x])dx_{n}$}
\end{prooftree}

\begin{prooftree}
\AxiomC{$M_0\twoheadrightarrow M_0',M_1\twoheadrightarrow M_1',M_2\twoheadrightarrow M_2',M_1,M_2:\basetype$}
\UnaryInfC{$\int_{M_1}^{M_2}\lambda y:T_2 .M_0dx~ \twoheadrightarrow \lambda y:T_2.\int_{M_1'}^{M_2'}M_0' dx$}
\end{prooftree}

\begin{prooftree}
\AxiomC{$N\twoheadrightarrow N',M\twoheadrightarrow M',M,N:\basetype,\forall j\in[1,n],M_j\twoheadrightarrow M_j'$}
\UnaryInfC{$\int_{M}^{N}(M_1,M_2,...M_{n})dx\twoheadrightarrow (\int_{M'}^{N'}M_1'dx,\int_{M'}^{N'}M_2'dx,...,\int_{M'}^{N'}M_{n}'dx)$}
\end{prooftree}

\begin{prooftree}
\AxiomC{$N\twoheadrightarrow N',M\twoheadrightarrow M',N:\basetype$}
\UnaryInfC{$(\lambda x:T .M) * N\twoheadrightarrow \lambda x:T .(M'*N')$}
\end{prooftree}

\begin{prooftree}
\AxiomC{$\forall j\in[1,n], M_j\twoheadrightarrow M_j',N\twoheadrightarrow N',N:\basetype$}
\UnaryInfC{$(M_1,M_2,...M_{n}) * N\twoheadrightarrow (M_1'*N',M_2'*N',...M_{n}'*N')$}
\end{prooftree}

\begin{prooftree}
\AxiomC{$\forall j\in[1,n], M_{1j}\twoheadrightarrow M_{1j}', M_{2j}\twoheadrightarrow M_{2j}'$}
\UnaryInfC{$(M_{11},M_{12},...M_{1n})*(M_{21},M_{22},...M_{2n})\twoheadrightarrow M_{11}'*M_{21}'\oplus M_{12}'*M_{22}'\oplus ... \oplus M_{1n}'*M_{2n}'$}
\end{prooftree}

\begin{prooftree}
\AxiomC{$M\twoheadrightarrow M'$}
\UnaryInfC{$fix~M\twoheadrightarrow M'~(fix~M'),fix~M\twoheadrightarrow fix~M'$}
\end{prooftree}

\begin{lemma}[Preservation of $\twoheadrightarrow$]
\xmodify{if}{If} $N:B,N\twoheadrightarrow N'$, then $N':B$.
\end{lemma}
\begin{proof}
\xmodify{if}{If} we name the one-step relation of our calculus as $\rho$, and its transitive closure as $\rho^*$, then we have $ \twoheadrightarrow \subseteq \rho^*$. So we have $N\rho^* N'$.  Notice that we have the preservation property of our calculus, thus we have $N':B$.

\end{proof}

\begin{lemma}[$\twoheadrightarrow$ under substitution]
$M\twoheadrightarrow M'$, $N\twoheadrightarrow N'$, then we have $M[N/x]\twoheadrightarrow M'[N'/x]$
\end{lemma}

\begin{proof}
Induction on $M\twoheadrightarrow M'$

\begin{itemize}
\item Case $M\twoheadrightarrow M$, make induction on the form of M.

\smallskip

\begin{itemize}

\item Subcase  c, \xmodify{Then}{then} c[N/x] = c = c[N'/x], using $ M\twoheadrightarrow M$ we have $M[N/x]\twoheadrightarrow M[N'/x]$.

\smallskip

\item Subcase $(t_1,t_2,...,t_{n})$, using induction we have $t_i[N/x]\twoheadrightarrow t_i[N'/x]$, Then using $ \forall i, M_i\twoheadrightarrow M_i' \Rightarrow (M_1,M_2,..,M_{n})\twoheadrightarrow (M_1',M_2',..,M_{n}')$ we have $ M\twoheadrightarrow M$ we have $M[N/x]\twoheadrightarrow M[N'/x]$.

\ Using the same technique, we can prove the subcase of $t\oplus t$, $t\ominus t$, $t * t$, $\lambda x:T. t$, $t\ t$
, $\myder{t}{x}{t}$, $\int_t^t tdx$, $\pi_j(t)$,  $\int_{M_1}^{M_2}M_3dx \twoheadrightarrow \int_{M_1'}^{M_2'}M_3'dx,inl/inr~M, case~inr~M_1~of~inl~x_1 \Rightarrow M_2|~inr~x_2 \Rightarrow M_3$.

\smallskip

\item Subcase variable y, if y = x then y[N/x] = N, y[N'/x] = N', then $ y[N/x]\twoheadrightarrow y[N'/x]$, if y is not x then same as the subcase c.

\end{itemize}

The rest cases can be divided into three categories.

\smallskip

\item Case relation based on the relation of subterms.

\begin{itemize}

\item Subcase $M~N\twoheadrightarrow M'~N'$, using induction we have $M[K/x]\twoheadrightarrow M'[K'/x]$, $N[K/x]\twoheadrightarrow N'[K'/x]$, using $ M\twoheadrightarrow M',N\twoheadrightarrow N' \Rightarrow M~N\twoheadrightarrow M'~N'$ we have $  (M~N)[K/x]\twoheadrightarrow (M'~N')[K'/x]$.

\smallskip

\item Subcases $M\oplus N\twoheadrightarrow M'\oplus N'$, $M\ominus N\twoheadrightarrow M'\ominus N'$, $M* N\twoheadrightarrow M'* N'$, $\myder{M}{x}{N}\twoheadrightarrow \myder{M'}{x}{N'}$, $(M_1,M_2,..,M_{n})\twoheadrightarrow (M_1',M_2',..,M_{n}')$, $\lambda x:T.M\twoheadrightarrow \lambda x:T.M',\pi_j(M)\twoheadrightarrow \pi_j(M')$, $fix~M\twoheadrightarrow fix~M'$, $inl/inr~M\twoheadrightarrow inl/inr~M'$: same as $M~N\twoheadrightarrow M'~N'$.

\end{itemize}

\smallskip

\item Case reduction changes the structure
\begin{itemize}

\item Subcase $ (M_{11},M_{12},...M_{1n}) \oplus (M_{21},M_{22},...M_{2n})  \twoheadrightarrow (M_{11}'\oplus M_{21}',M_{12}'\oplus M_{22}',...M_{1n}'\oplus M_{2n}')$, using induction we have $\forall i\in[1,2],\forall j \in[1,n],M_{ij}[K/x]\twoheadrightarrow  M_{ij}'[K'/x]$, so we have

 $ ((M_{11},M_{12},...M_{1n}) \oplus (M_{21},M_{22},...M_{2n}))[K/x] \twoheadrightarrow (M_{11}'\oplus M_{21}',M_{12}'\oplus M_{22}',...M_{1n}'\oplus M_{2n}')[K'/x]$.

\smallskip

\item Subcases  $(M_{11},M_{12},...M_{1n}) \ominus (M_{21},M_{22},...M_{2n})  \twoheadrightarrow (M_{11}'\ominus M_{21}',M_{12}'\ominus M_{22}',...M_{1n}'\ominus M_{2n}')$, $\myder{(M_1,M_2,..,M_{n})}{x}{N}\twoheadrightarrow (\myder{M_1'}{x}{N'},\myder{M_2'}{x}{N'},...,\myder{M_{n}'}{x}{N'})$, $\myder{\lambda y:T. M}{x}{N}\twoheadrightarrow\lambda y:T.\myder{M'}{x}{N'}$, $\int_{M_1}^{M_2}\lambda y:T_2 .M_0dx~ \twoheadrightarrow \lambda y:T_2.\int_{M_1'}^{M_2'}M_0' dx$, $\int_{M}^{N}(M_1,M_2,...M_{n})dx\twoheadrightarrow (\int_{M'}^{N'}M_1'dx,\int_{M'}^{N'}M_2'dx,...,\int_{M'}^{N'}M_{n}'dx)$, $(\lambda x:T .M) * N\twoheadrightarrow \lambda x:T .(M'*N')$, $(M_1,M_2,...M_{n}) * N\twoheadrightarrow (M_1'*N',M_2'*N',...M_{n}'*N')$, $(M_{11},M_{12},...M_{1n})*(M_{21},M_{22},...M_{2n})\twoheadrightarrow M_{11}'*M_{12}'\oplus M_{21}'*M_{22}'\oplus ... \oplus M_{1n}'*M_{2n}'$, $fix~M\twoheadrightarrow M'~(fix~M')$: same as $ (M_{11},M_{12},...M_{1n}) \oplus (M_{21},M_{22},...M_{2n})  \twoheadrightarrow (M_{11}'\oplus M_{21}',M_{12}'\oplus M_{22}',...M_{1n}'\oplus M_{2n}')$.

\end{itemize}

\smallskip

\item Case reduction involves substitution

\begin{itemize}

\item Subcase $(\lambda x:T.M)N \twoheadrightarrow M'[N'/x]$, by induction hypothesis, we have $M[K/y]\twoheadrightarrow M'[K'/y]$, $N[K/y]\twoheadrightarrow N'[K'/y]$, thus $((\lambda x:T.M)N)[K/y]= ((\lambda x:T.M[K/y])N[K/y]) \twoheadrightarrow M'[K'/y]([(N'[K'/y])/x)$, and we have $  M'[N'/x][K'/y]    =M'[K'/y]([(N'[K'/y])/x)$, Therefore we prove the case.

\smallskip

\item Subcase$ \forall i,(M_1', M_2'...,M_{i-1}',x_i,M_{i+1}'...,M_{n}')is~written~as~ M_{i*}'
,\myder{ M_0}{ x}{(M_1, M_2,...,M_{n})} \twoheadrightarrow$

$ (\myder{M_0'[M_{1*}'/x]}{x_1}{M_1'},\myder{ M_0[M_{2*}'/x]}{x_2}{M_2'},...,\myder{t[M_{n*}'/x]}{x_{n}}{M_{n}'})$

Notice that 
\xmodify{put the equation into array}{}
\[
\begin{array}{llll}\\
M_0'[M_{i*}'/x][K'/y]\\
 \quad = \quad M_0'[K'/y][(M_{i*}'[K'/y])/x]\\
					  \quad	= \quad (M_0'[K'/y])[(M_1'[K'/y], M_2'[K'/y],...,\\
\phantom{\quad	= \quad (M_0'[K'/y])[(}M_{i-1}'[K'/y], x_i[K'/y],M_{i+1}'[K'/y],...,M_{n}'[K'/y])/x]\\
\quad =  \quad(M_0'[K'/y])[((M'[K'/y])_{i*}')/x]

\end{array}
\] Using induction, we know that $M_i[K/y] \twoheadrightarrow M_i'[K'/y]$, so we prove the case.

\smallskip

\item Subcase $(\lambda x:T. M )\oplus (\lambda y:T. N) \twoheadrightarrow \lambda x:T. M' \oplus N'[y/x]$, $(\lambda x:T. M )\ominus (\lambda y:T. N) \twoheadrightarrow \lambda x:T. M' \ominus N'[y/x]$: same as $(\lambda x:T.M)N \twoheadrightarrow M'[N'/x]$.

\smallskip

\item Subcases $\int_{(M_{11},M_{12},...M_{1n}) }^{(M_{21},M_{22},...,M_{2n})}M_0dx~ \twoheadrightarrow \int_{M_{11}'}^{M_{21}'}  \pi_1(M_0'[M_{1*}'/x])dx_1 \oplus ... \oplus \int_{M_{1n}'}^{M_{2n}'} \pi_n(M_0'[M_{n*}'/x])dx_{n}$,
$case~inl~M_1~of~inl~x_1 \Rightarrow M_2|~inr~x_2 \Rightarrow M_3 \twoheadrightarrow M_2'[M_1'/x_1]$
,$case~inr~M_1~of~inl~x_1 \Rightarrow M_2|~inr~x_2 \Rightarrow M_3 \twoheadrightarrow M_3'[M_1'/x_2]$
: same as $\forall i,(M_1', M_2',M_{i-1}',x_i,M_{i+1}'...,M_{n}')$is written as  $M_{i*}'$
,$\myder{ M_0}{ x}{(M_1, M_2,...,M_{n})} \twoheadrightarrow$
$ (\myder{M_0'[M_{1*}'/x]}{x_1}{M_1'},\myder{ M_0[M_{2*}'/x]}{x_2}{M_2'},...,\myder{t[M_{n*}'/x]}{x_{n}}{M_{n}'})$.

\end{itemize}

Thus we complete the proof.

\end{itemize}

\end{proof}

\begin{lemma}[diamond property]
\xmodify{for}{For} $M\twoheadrightarrow M_1,M\twoheadrightarrow M_2$, there exists a $M_3$, \xmodify{that}{such that} $M_1\twoheadrightarrow M_3,M_2\twoheadrightarrow M_3$
\end{lemma}

\begin{proof}

\xmodify{we make}{We do} induction on the case of $M\twoheadrightarrow M_1$.

\begin{itemize}

\item Case $ M\twoheadrightarrow M$ \xmodify{Change it into the next line}{}

Then we choose $M_3$ as $M_2$.

\smallskip

\item Case $ M~N\twoheadrightarrow M'~N'$

\begin{itemize}

\item Subcase $M\twoheadrightarrow M_2$ as $ M\twoheadrightarrow M$

Then we choose $M_3$ as $M_1$.

\smallskip

\item Subcase $M\twoheadrightarrow M_2$ as $  M~N\twoheadrightarrow M''~N''$ \textcircled{1}

Then we use \xmodify{induction}{the induction} hypothesis, we have $M^*$ that $M'\twoheadrightarrow M^*$,$M''\twoheadrightarrow M^*$,\xmodify{we have}{and we have} $N^*$ that $N'\twoheadrightarrow N^*$,$N''\twoheadrightarrow N^*$, so we choose $M_3$ as $M^*~N^*$.

\smallskip

\item Subcase $M\twoheadrightarrow M_2$ as $M = \lambda x:T.P$, $ (\lambda x:T.P)N \twoheadrightarrow P''[N''/x]$. \textcircled{2}

Then we first have that $M = \lambda x:T.P$ Then $M' = \lambda x:T.P'$, \xmodify{So}{so} we choose $M_3=P^*[N^*/x]$.

\end{itemize}

\smallskip

\item Case $M\oplus N\twoheadrightarrow M'\oplus N'$

\begin{itemize}

\item Subcase $M\twoheadrightarrow M_2$ as $ M\twoheadrightarrow M$

Then we choose $M_3$ as $M_1$.

\smallskip

\item Subcase $M\twoheadrightarrow M_2$ as $  M\twoheadrightarrow M'',N\twoheadrightarrow N'' \Rightarrow M\oplus N\twoheadrightarrow M''\oplus N''$: same as \textcircled{1}.

\smallskip

\item Subcase $M\twoheadrightarrow M_2$ as $(\lambda x:T. M )\oplus (\lambda y:T. N) \twoheadrightarrow \lambda x:T. M' \oplus N'[y/x]$ \xmodify{move to the next line}{}

 Then $M\twoheadrightarrow M_1$ must be $(\lambda x:T. M )\oplus (\lambda y:T. N)\twoheadrightarrow  (\lambda x:T. M'' )\oplus (\lambda y:T. N'')$, Then we choose $M_3$ to be $\lambda x:T. M^* \oplus N^*[y/x]$.

\smallskip

\item Subcase $M\twoheadrightarrow M_2$ as $(M_{11},M_{12},...M_{1n}) \oplus (M_{21},M_{22},...M_{2n})  \twoheadrightarrow (M_{11}'\oplus M_{21}',M_{12}'\oplus M_{22}',...M_{1n}'\oplus M_{2n}')$: same as $(\lambda x:T. M )\oplus (\lambda y:T. N) \twoheadrightarrow \lambda x:T. M' \oplus N'[y/x]$.

\end{itemize}

All the other cases are \xmodify{similar}{similar to} the case of application and $\oplus$, except that we may have more subcases on these cases, but the extra subcases are all similar to \textcircled{2}.

\end{itemize}
\end{proof}

\begin{lemma}[Confluence]\label{ConfluenceProof}
One term has at most one normal form.
\end{lemma}
\begin{proof}
The relation $\twoheadrightarrow$ has the diamond property, and reduction relation $\rho$ satisfy that $\rho \subseteq \twoheadrightarrow \subseteq \rho^*$\xmodify{,}{.} Also notice that $\twoheadrightarrow^*$ has the diamond property, and $\twoheadrightarrow^* = \rho^*$, so the relation $\rho^*$ has the diamond property\xmodify{, There comes}{. This is what results in} the confluence.
\end{proof}

\subsection{Strong normalization}

Here we write $\rightsquigarrow$ as $\rho^*$.
\xmodify{switch the following t and t'}{$t$ and $t'$}
\begin{lemma}[existence of $\nu$]\rm$t$ is strongly normalisable iff there is a number $\nu(t)$ which bounds the
length of every normalisation sequence beginning with $t$. 
\end{lemma}
\xmodify{none}{
\begin{proof}
See P27 in Proofs and Types \cite{Gir89}.
\end{proof}
}
\begin{definition}\rm
We define a set $\textbf{RED}_T$ by induction on the type $T$.

1. For $t$ of base type, $t$ is reducible iff it is strongly normalisable.

2. For $t$ of type $(T_1,T_2,...,T_{n})$ , $t$ is reducible iff $\forall j,\pi_j(t)$ is reducible.

3. For $t$ of type U$\rightarrow$V , $t$ is reducible iff, for all reducible $u$ of type $U$, $t~u$ is
reducible of type V .

4. For $t$ of type $T_1+ T_2$, $t$ is reducible iff, $case~t~of~inl~x_1\Rightarrow 0~|~inr~x_2\Rightarrow 0$is reducible term of base type.
\end{definition}

\begin{definition}
\rm
$t$ is neutral if $t$ is not of the form $(t_1,t_2,...,t_{n})$ or $\lambda x:T .t$ or $inl/inr~t$

\end{definition}

We \xmodify{would}{will} verify the following 3 properties by induction on types.

(CR 1) If $t$ $\in$ $\textbf{RED}_T$ , then t is strongly normalisable.

(CR 2) If $t$ $\in$ $\textbf{RED}_T$ and $t$$\rightsquigarrow$$t'$, then $t'$ $\in$ $\textbf{RED}_T$ .

(CR 3) If $t$ is neutral, and whenever we convert a redex of $t$ we obtain a term
t'$\in$ $\textbf{RED}_T$ , then t$\in$ $\textbf{RED}_T$ .

\begin{itemize}

\item Case base type

(CR 1) is a tautology.

(CR 2) If $t$ is strongly normalisable then every term \xmodify{t'to}{$t'$ to} which $t$ reduces is also.

(CR 3) A reduction path leaving $t$ must pass through one of the terms $t'$
, which
are strongly normalisable, and so is finite. In fact, it is immediate that $\nu$(t) is equal to the greatest of the numbers $\nu(t')+1$, as $t'$ varies over the (one-step) conversions of $t$.

\smallskip

\item Case tuple type

(CR 1) Suppose that $t$, of type $(T_1,T_2,...,T_n)$ , is reducible; then $\pi_1(t)$ is reducible and by induction hypothesis (CR 1) for $T_1$, $\pi_1(t)$ is strongly normalisable. Moreover, $\nu(t) \le \nu(\pi_1(t))$. \xmodify{since}{Since} to any reduction sequence $t$, $t_1$, $t_2$, . . ., one can apply $\pi_1()$
to construct a reduction sequence $\pi_1(t)$, $\pi_1(t_1)$, $\pi_1(t_2)$... (in which the $\pi_1()$ is not reduced). So $\nu(t)$ is finite, and t is strongly normalisable.

(CR 2) If $t\rightsquigarrow t'$, then $\forall j,\pi_j(t)\rightsquigarrow \pi_j(t')$. \xmodify{induction}{By the induction} hypothesis for type $T_j$ on (CR 2), we have $\forall j,\pi_j(t')$ is reducible, so $t'$ is reducible

(CR 3) Let $t$ be neutral and suppose all the $t'$ one step from t are reducible.
Applying a conversion inside $\pi_j(t)$, the result is a $\pi_j(t')$
, since $\pi_j(t)$ cannot itself be a redex ($t$ is not a tuple), and $\pi_j(t')$ is reducible, since $t'$ is. But as $\pi_j(t)$ is neutral, and all the terms one step from $\pi_j(t)$ are reducible, the induction hypothesis (CR 3) for $T_j$ ensures that $\pi_j(t)$ is reducible. so
$t$ is reducible.

\smallskip

\item Case arrow type

(CR 1) If $t$ is reducible of type $U\rightarrow V$ , let $x$ be a variable of type $U$\modify{;}{.} And we have $\nu(t) \le \nu(t~x)$

(CR 2) If $t\rightsquigarrow t'$, and t is reducible, take u reducible of type $U$; then $t~u$ is
reducible and $t~u\rightsquigarrow t'~u$ The induction hypothesis (CR 2) for $V$ gives that
$t'~ u$ is reducible. So $t'$ is reducible.

(CR 3) Let $t$ be neutral and suppose all the $t'$ one step from $t$ are reducible. Let $u$ be a reducible term of type $U$; we want to show that $t~ u$ is reducible. By induction hypothesis (CR 1) for $U$, we know that $u$ is strongly normalisable; so we can reason by induction on $\nu(u)$.

In one step, $t~u$ converts to
\begin{itemize}

\item 1. $t~ u$ with $t'$ one step from $t$; but $t'$ is reducible, so $t~ u$ is.

\smallskip

\item 2. $t~u'$, with $u'$ one step from $u$. $u'$ is reducible by induction hypothesis(CR 2) for U, and $\nu(u'
) < \nu(u)$; so the induction hypothesis for $u'$ tells us that $t~u'$ is reducible.

\smallskip

\item 3. There is no other possibility, for $t~u$ cannot itself be a redex ($t$ is not
of the form $\lambda x:T. t$).

\end{itemize}
\item Case sum type

(CR 1) If $t$ is reducible of type $T_1+T_2$ , Then we have $\nu(t) \le \nu(case~t~of~inl~x_1\Rightarrow 0|inr~x_2\Rightarrow 0)$

(CR 2) same as tuple type.

(CR 3) same as arrow type.

\end{itemize}

\begin{lemma}\rm
\xmodify{if}{If} $t_1,t_2,...,t_{n}$ are reducible terms, then \xmodify{so $(t_1,t_2,...,t_{n})$ is}{so is $(t_1,t_2,...,t_{n})$}
\end{lemma}

\begin{proof}

Because of (CR 1), we can reason by induction on $\nu(t_1) + \nu(t_2)+...+\nu(t_{n})$ to show that $\pi_j(t_1,t_2,...,t_{n})$, is reducible. This term converts to

\begin{itemize}
\item 1. $t_j$, then it is reducible.

\smallskip

\item 2.$(t_1,...,t_{k-1},t_k',t_{k+1},...,t_{n})$, based on induction, it is reducible.

\end{itemize}

\end{proof}

\begin{lemma}\rm
\xmodify{if}{If} for all reducible u of type U, t[u/x] is reducible, then so is $\lambda x:T. t$.
\end{lemma}

\begin{proof}

To show $\lambda x:T.t~u$ is reducible, \xmodify{We}{we} make reductions on $\nu(u) + \nu(t)$, $\lambda x:T.t~u$ \xmodify{can}{, which can} be reduced to

\begin{itemize}
\item 1. $t[u/x]$, then it is reducible.

\smallskip

\item 2. $(\lambda x:T.t')~u$ or  $(\lambda x:T.t)~u'$, based on induction we know it is reducible.

\end{itemize}
\end{proof}

\begin{lemma}\rm
\xmodify{if}{If} $t$ is reducible, then so is $inl/inr~t$.
\end{lemma}

\begin{proof}
\xmodify{same}{Same} as the case $\lambda x:T. t$.
\end{proof}

\begin{lemma}\rm
\xmodify{if}{If} for all reducible $t_1$ and $t_2$ of type $T_1$ and $T_2$, we have $t_3[t_1/x_1]$ and $t_4[t_2/x_2]$ are reducible, and t is reducible term of type $T_1+T_2$, then so is $case~t~of~inl~x_1\Rightarrow t_3~|~inr~x_2\Rightarrow t_4$.
\end{lemma}

\begin{proof}
\xmodify{same}{Same} as the case $\lambda x:T. t$.
\end{proof}

\begin{lemma}\rm
\xmodify{if}{If} $t_1$ and $t_2$ are reducible terms of T, then so is $t_1 \oplus t_2$.
\end{lemma}

\begin{proof}

We prove this by induction on type.

\begin{itemize}
\item Case base type, then it can only be reduced to $t_1'\oplus t_2'$, so $\nu(t_1\oplus t_2) =\nu(t_1) + \nu(t_2)$, Therefore it is  strongly normalisable, and thus reducible.

\smallskip

\item Case $(T_1,T_2,...,T_{n})$, we make induction on $\nu(t_1) + \nu(t_2)$, $t_1\oplus t_2$ can be reduced to.

\begin{itemize}

\item 1.Subcase $(t_{11},t_{12},...t_{1n}) \oplus (t_{21},t_{22},...t_{2n})  \rightarrow (t_{11}\oplus t_{21},t_{12}\oplus t_{22},...t_{1n}\oplus t_{2n})$. Because $t_1$ and $t_2$ si reducible, then $\forall i\forall j,t_{ij}$ is reducible, based on induction on types, we have $\forall j, t_{1j}\oplus t_{2j}$ Thus, $(t_{11}\oplus t_{21},t_{12}\oplus t_{22},...t_{1n}\oplus t_{2n})$ is reducible.

\smallskip

\item 2.Subcase $(t_1'\oplus t_2)$ or $(t_1\oplus t_2')$. Based on induction, we know it is reducible.

\end{itemize}
\item Case $A \rightarrow B$ for all reducible term u of type A, we make induction on $\nu(t_1) + \nu(t_2) + \nu(u)$.

\begin{itemize}
\item 1. Subcase $(\lambda x: T.t_1) \oplus (\lambda y: T.t_2)  \rightarrow \lambda x: T.t_1 \oplus  (t_2[x/y])$ notice that for all reducible u of type A $(t_1 \oplus t_2[x/y])[u/x]$, notice that this term is equal to $(t_1[u/x] \oplus t_2[u/y])$, because $(\lambda x: T.t_1)$ is a reducible term, then so is $t_1[u/x]$. Because $t_1[u/x]$ and $t_2[u/y]$ are reducible terms based on induction. So we have $\lambda x: T.t_1 \oplus  (t_2[x/y])$'s reducibility.

\smallskip

\item 2 Subcase  $(t_1' \oplus t_2)~u$ or $(t_1 \oplus t_2')~u$ or $(t_1 \oplus t_2)~u'$, based on induction we can prove the case.

\end{itemize}
\end{itemize}
\end{proof}

\begin{lemma}\rm
\xmodify{if}{If} $t_1$ and $t_2$ are reducible terms of T, then so is $t_1 \ominus t_2$.

\end{lemma}

\begin{proof}
Same as $\oplus$.
\end{proof}

\begin{lemma}\rm
\xmodify{if}{If} $t_1$ and $t_2$ are reducible terms of $\frac{\partial T_1}{\partial T_2}$ and $T_2$, then so is $t_1 * t_2$.

\end{lemma}

\begin{proof}
We prove this by induction on \xmodify{type}{types}.

\begin{itemize}
\item Case $T_1$: base type, $T_2$: base type: same as the case of $\oplus$.

\smallskip

\item Case $T_1$: $(T_1,T_2,...,T_{n})$, $T_2$: base type: same as the case of $\oplus$.

\smallskip

\item Case $T_1$: $A\rightarrow B$, $T_2$: base type: same as the case of $\oplus$.

\smallskip

\item Case $T_1$: $(T_1,T_2,...,T_{n})$, $T_2$: $(T_1',T_2',...,T_{n}')$

Suppose $ t_1:(t_{11},t_{12},...t_{1n}),t_2 :(t_{21},t_{22},...t_{2n}) $, we \xmodify{make}{do} induction on $\nu(t_1)+\nu(t_2)$.

\begin{itemize}
\item Subcase $t_1 * t_2\rightarrow (t_{11}*t_{21})\oplus (t_{12}*t_{22})\oplus ... \oplus (t_{1n}*t_{2n}) $: Because \xmodify{$t_1$, $t_2$}{$t_1$ and $t_2$} is reducible, then so is $\forall i\forall j, t_{ij}$, based on induction on types we have $\forall j,t_{1j}*t_{2j}$ is reducible, then so is $(t_{11}*t_{21})\oplus (t_{12}*t_{22})\oplus ... \oplus (t_{1n}*t_{2n})$.

\smallskip

\item Subcase $t_1' *t_2$ or $t_1 * t_2'$, based on induction we know it is reducible.

\end{itemize}

\end{itemize}

\end{proof}

\begin{lemma}\rm
\xmodify{if}{If} $t_1$ and $t_2$ are reducible terms of $T_1$ and $T_2$, and for all reducible u of type $T_2$, we \xmodify{have}{have that} $t_1[u/x]$ is reducible then so is $\myder{t_1}{x}{t_2}$.
\end{lemma}

\begin{proof}
we prove this by induction on \xmodify{type}{types}.

\begin{itemize}
\item Case $T_1$: $(T_1,T_2,...,T_{n})$ or $A\rightarrow B$ or base type, $T_2$:\basetype. Same as the case \xmodify{*}{$t_1~*~t_2$}.

\smallskip

\item Case $T_1$: $(T_1,T_2,...,T_{n})$, $T_2$: $(T_1',T_2',...,T_{n}')$, we make induction on $\nu(t_1)+\nu(t_2)$.

\begin{itemize}
\item Subcase $\forall i,(t_1, t_2...,t_{i-1},x_i,t_{i+1}...,t_{n})is~written~as~ t_{i*}$,

$\myder{t}{x}{(t_1, t_2,...,t_{n})} \rightarrow (\myder{t[t_{1*}/x]}{x_1}{t_1},\myder{t[t_{2*}/x]}{x_2}{t_2},...,\myder{t[t_{n*}/x]}{x_{n}}{t_{n}})$, note that $t_{i*}$ is reducible so based on \xmodify{induction we have}{induction. We have that} $\myder{t[t_{j*}/x]}{x_j}{t_j}$ is reducible. Note that this induction is based on the hypothesis $(t[t_{j*}/x])[u/x_j]$ is reducible for all the reducible u of type $T_j'$, and  $(t[t_{j*}/x])[u/x_j] = (t[(t_{j*}[u/x_j])/x])$ because t has no occurrence of $x_j$, and it is easy to show that $(t_{j*}[u/x_j])$ is a reducible term of type $T_2$, so we finish the induction, then we have

$(\myder{t[t_{1*}/x]}{x_1}{t_1},\myder{t[t_{2*}/x]}{x_2}{t_2},...,\myder{t[t_{n*}/x]}{x_{n}}{t_{n}})$ is reducible.

\smallskip

\item Subcase $\myder{t_1'}{x}{t_2}$ or $\myder{t_1}{x}{t_2'}$, based on induction we have the proof.

\end{itemize}
\end{itemize}
\end{proof}

\begin{lemma}\rm \xmodify{if}{If} $t_1$,$t_2$ and $t_3$ are reducible terms of $T_1$, $T_1$ and $T_2$, and for all reducible u of type $T_1$, we have $t_3[u/x]$ is reducible then so is $\int_{t_1}^{t_2}t_3dx$.
\end{lemma}

\begin{proof}
Same as the case of $\myder{}{ x}{...}$
\end{proof}

\begin{lemma}\rm \xmodify{if}{If} $t_1$,$t_2$ and $t_3$ are reducible terms of $T_1+T_2$, $T$ and $T$, and for all reducible $u_1$ of type $T_1$, $u_2$ of type $T_2$, we \xmodify{have}{have that} $t_2[u_1/x_1]$ and $t_3[u_2/x_2]$ are reducible then so is $\mycase{t_1}{x_1}{t_2}{x_2}{t_3}$.
\end{lemma}

\begin{proof}
Same as the case of $\myder{}{ x}{...}$
\end{proof}

\begin{lemma}\rm\label{ReducibleLemma} Let $t$ be any term (not assumed to be reducible), and suppose all
the free variables of t are among $x_1,...,x_n$ of types $U_1, . . . , U_n$. If $u_1,...,u_n$ are
reducible terms of types $U_1 ,..., U_n$ then $t[u_1/x_1,..., u_n/x_n]$ is reducible.
\end{lemma}

\begin{proof} By induction on $t$. We write t[\underline{u}/\underline{x}] for $t[u_1/x_1,..., u_n/x_n]$.

\begin{itemize}
\item 1. $t$ is $x_i$, then \xmodify{it $t[u/x_i]$}{$t[u/x_i]$} is reducible.

\smallskip

\item 2. $t$ is $c$, then $t_i$ has no free variable, and $c$ itself is reducible, so it is reducible.

\smallskip

\item 3. $t$ is $(t_1,t_2,...,t_n)$, based on induction we prove $t_j$[\underline{u}/\underline{x}] is reducible, based on the lemma we know it is reducible.

\smallskip

\item 4. $t$ is $t \oplus t$, $t \ominus t$, $t * t$,$inl/inr~t$ or $\pi_j(t)$: same as the case $(t_1,t_2,...,t_n)$.

\smallskip

\item 5. $t$ is $\lambda y:T .t$, by induction we have t[\underline{u}/\underline{x},v/y] is reducible, then by lemma we have  $\lambda y:T .(t[\underline{u}/\underline{x}])$ is reducible, so $(\lambda y:T .t)[\underline{u}/\underline{x}]$ is reducible.

\smallskip

\item 6. $t$ is $\myder{t}{x}{t}$, $\mycase{t}{x_1}{t_1}{x_2}{t_2}$ or $\int_t^ttdx$: same as the case $\lambda y:T .t$.

\end{itemize}
\end{proof}
\begin{theorem}
\rm

All terms are reducible.

\end{theorem}

\xmodify{none}{
\begin{proof}
For arbitrary term $t$, apply the Lemma \ref{ReducibleLemma} to  $t[x_1/x_1,..., x_n/x_n]$ and we get the result.
\end{proof}
}
\begin{corollary}\label{NormalizationProof}

\rm

All terms are strongly normalisable.

\end{corollary}
\section{Appendix-Lemmas}
\begin{lemma}
  \rm
\xmodify{if}{If} $t_1 \rho^* t_1',t_2 \rho^* t_2'$,~\xmodify{none}{then}~$t_1[t_2/x] \rho^* t_1'[t_2'/x]$.
\end{lemma}
\begin{proof}
Using the confluence property, it is easy to see.
\end{proof}

\begin{lemma}
\rm
\xmodify{if}{If}  $t_1 =t_1'$, $t_2=t_2'$, then $t_1\oplus t_2 = t_1'\oplus t_2' $.
\end{lemma}

\begin{proof}

if $t_1$ or $t_2$ is not closed, then we use the substitution $[u_1/x_1,..., u_n/x_n]$ to make it closed. For simplicity of notation, we just use $t_1$ and $t_2$ to be the closed-term of themselves.

Based on the equality defintion, we can assume that $t_1$, $t_2$, $t_1'$, $t_2'$ are all normal forms and we prove this by induction on \xmodify{type}{types}.

\begin{itemize}

\item Case $t_1$ is of base type. \xmodify{then}{Then} $t_2$, $t_1'$ and $t_2'$ have to be base type to be well-typed. \xmodify{and}{And} for base type normal forms, we have $t_1\oplus t_2 = t_1'\oplus t_2' $.

\smallskip

\item Case $t_1$ is $A\rightarrow B$ type, let's suppose $t_1:\lambda x:T.t_3$, $t_1':\lambda x':T.t_3'$, $t_2:\lambda y:T.t_4$, $t_2':\lambda y':T.t_4'$.

\xmodify{if}{If} $t_1$ or $t_2$ or $t_1'$ or $t_2'$ 's normal form are not $\lambda x:T.t$, then we know their normal form are all interpretable in base type, thus we have $t_1\oplus t_2 = t_1'\oplus t_2' $.

 Else for all u
 \[
\begin{array}{llll}
(t_1\oplus t_2)~u\\
\qquad = (\lambda x:T.t_3\oplus \lambda y:T.t_4)~u\\
\qquad = (\lambda x:T.t_3\oplus t_4[x/y])~u\\
\qquad = t_3[u/x]\oplus t_4[u/y]
\end{array}
\]

Similarly, we have $(t_1'\oplus t_2')~u' =  t_3'[u/x']\oplus t_4'[u/y']$.

And notice that because $t_1=t_1'$, so $t_1~u=t_1'~u$, so $t_3[u/x] = t_3'[u/x']$, based on induction of type B, we have $ t_3[u/x]\oplus t_4[u/y] =  t_3'[u/x']\oplus t_4'[u/y']$, \xmodify{so we prove}{so we have proven} the case.

\smallskip

\item Case $t_1$ is of type $(T_1,T_2,..,T_{n})$. Then we suppose $t_1:(t_{11},t_{12},..,t_{1n})$, $t_1':(t_{11}',t_{12}',..,t_{1n}')$, $t_2:(t_{21},t_{22},..,t_{2n})$, $t_2':(t_{21}',t_{22}',..,t_{2n}')$

Then
\[
\begin{array}{llll}
t_1 \oplus t_2\\
= (t_{11},t_{12},..,t_{1n})\oplus (t_{21},t_{22},..,t_{2n})\\
=(t_{11}\oplus t_{21},t_{12}\oplus t_{22},..,t_{1n}\oplus t_{2n})
\end{array}
\]

Similarly we have $t_1' \oplus t_2' = (t_{11}'\oplus t_{21}',t_{12}'\oplus t_{22}',..,t_{1n}'\oplus t_{2n}')$, and based on induction we have $\forall j,t_{1j}\oplus t_{2j} =t_{1j}'\oplus t_{2j}'$, so we have $t_1 \oplus t_2$ = $t_1' \oplus t_2'$.

\item Case $t_1$ is of type $T_1+T_2$. This case is impossible because it is not well-typed.

\end{itemize}
\end{proof}

\begin{lemma}
\rm
For a term t, for any subterm s, if the term s'=s, then t[s'/s]=t. (We only substitute the subterm s, but not other subterms same as s)

\end{lemma}

\begin{proof}

We \xmodify{prove by induction, we}{prove this by induction. We}  first substitute for all the free variables in \xmodify{t. then}{$t$. Then}

\[
\begin{array}{llll}

 t[s'/s][u_1/x_1,..., u_n/x_n]\\

 \qquad= t[u_1/x_1,..., u_n/x_n][s'[u_1/x_1,..., u_n/x_n]/s[u_1/x_1,..., u_n/x_n]]

 \end{array}
\]
 notice that $s'[u_1/x_1,..., u_n/x_n]=s[u_1/x_1,..., u_n/x_n]$ because $s'=s$, and we just substitute for some of the free variables in $s'$ and $s$. So we only need to prove that for a closed-term $t$, for any subterm $s$, if the term $s'=s$, then $t[s'/s]=t$.

And notice that if we choose the subterm $s$ to be the $t$ itself, then we have $t[s'/s]=s'=s=t$, And we prove the case. So we next make induction on the form of $t$.
\begin{itemize}

\item Case $t$ is $(t_1,t_2,...,t_{n})$

\ \ Using induction, we know $t_i[s'/s]=t_i$, and we want to prove
\[
\begin{array}{llll}
(t_1,t_2,...,t_{n})\\
\qquad=(t_1[s'/s],t_2[s'/s],...,t_{n}[s'/s])
\end{array}
\]

 And because we have the transitive property of equality, then we just reduce both of them to normal forms, then by definition we know they equal to each other, thus we prove the case.

\ \ Using the same technique, we can prove the case $\lambda x:T. t$ and $inl/int~t$.

\smallskip

\item Case $t$ is $c$

\ \ Then it has no subterm except $c$ itself, if $s'=c$, then $c[s'/c]=s'$, thus we prove the case.

\ \ Using the same technique, we can prove the case $x$.

\smallskip

\item Case $t$ is $t_1\oplus t_2$

\ \ Using induction we have $t_1[s'/s]=t_1$,$t_2[s'/s]=t_2$, and we have \xmodify{proved the}{proven} Lemma \ref{EqAdd} that if $t_1 =t_1',t_2=t_2'$, then $t_1\oplus t_2 = t_1'\oplus t_2' $, thus we \xmodify{prove}{have proven} the case.

\ \ Using the same technique, we can prove the case $t\ominus t$, $t*t$,  $\pi_j(t)$.

\smallskip

\item Case $t$ is $t_1~t_2$

\ \ We want to prove if $t_1=t_1',t_2=t_2'$, then $t_1~t_2=t_1'~t_2'$.

\ \ By definition we know if $t_1=t_1'$, then $t_1~t_2'=t_1'~t_2'$.

\ \ Then \xmodify{we prove $t_1~t_2'=t_1~t_2$,}{we have proven $t_1~t_2'=t_1~t_2$.}  Using confluence property, we can reduce the $t_1$ to $\lambda x:A.t$ or a \textbf{nb}. If it is the former case, then we using induction we have $t[t_2'/x]=t[t_2/x]$ .Thus we have  $t_1~t_2'=t_1~t_2$. If it is the latter case, then $t_2$'s normal form can be is interpretable, and on base type interpretation we have if $x=x'$, then $f(x)=f(x')$. Thus we prove the case.

\smallskip

\item Case $t$ is $\myder{t_1}{x}{t_2}$

\ \ \xmodify{if}{If} $t_2$ is base type, then we can use the \xmodify{same technique of that we prove the case}{techniques for how we proved the case } $t_1*t_2$,

\ \ \xmodify{if}{If} $t_2$ is of type $(T_1,T_2,...,T_{n})$, we can reduce the $t_2$ and $t_2'$ to the normal forms $(t_1, t_2,...,t_{n})$ and $(t_1', t_2',...,t_{n}')$, and then we have \reductiong, using induction we have $t[t_{j*}/x]=t[t_{j*}'/x]$, then based on induction we have $\myder{t[t_{j*}/x]}{x_j}{t_j}=\myder{t[t_{j*}'/x]}{x_j'}{t_j'}$, Thus we \xmodify{prove}{have proven} the case.

\ \ Using the same technique, we can prove the case $\int_{t_1}^{t_2}t_3dx$ adn $\mycase{t}{x_1}{t_1}{x_2}{t_2}$.

\end{itemize}

Thus we \xmodify{prove}{have proven} the lemma.

\end{proof}

\begin{lemma}
\rm
If $t_1*(t_2\oplus t_3)$ and $(t_1*t_2)\oplus (t_1*t_3)$ are weak-normalizable, then $t_1*(t_2\oplus t_3) = (t_1*t_2)\oplus (t_1*t_3)$
\end{lemma}
\begin{proof}

if $t_1$, $t_2$ and $t_3$ are not closed, then we use the substitution $[u_1/x_1,..., u_n/x_n]$ to make it closed. For simplicity of notation, we just use $t_1$, $t_2$ and $t_3$ to be the closed-term of themselves.

Because of the confluence and normalization property of the system, we can assume that $t_1$, $t_2$, $t_3$ are all normal formss and we prove this by induction on type.

\begin{itemize}

\item Case:$t_1$, $t_2$ and $t_3$ are of base type. then based on base type interpretation, we have $t_1*(t_2\oplus t_3) = (t_1*t_2)\oplus (t_1*t_3)$.

\smallskip

\item Case:$t_1$ is of type $A\rightarrow B$, $t_2$ and $t_3$ are of base type. Suppose $t_1$ is $\lambda x:A.t$.

\ \ If $t_1$ 's normal form is not $\lambda x:A.t$, then we notice that $t_1=\lambda x:A.t_1~x$, use the Lemma \ref{EqLSub}, we know we can use $\lambda x:A.t_1~x$ to substitute for $t_1$, Thus we can suppose $\lambda x:A.t$.

\ \ Then we have for all u of type A,
\[
\begin{array}{llll}
 t_1*(t_2\oplus t_3)~u\\
\qquad = (\lambda x:A.t)*(t_2\oplus t_3)~u\\
\qquad = \lambda x:A.(t*(t_2\oplus t_3))~u \\
\qquad= t[u/x]*(t_2\oplus t_3)
\end{array}
\]

\ \ And

\[
\begin{array}{llll}
 (t_1*t_2)\oplus (t_1*t_3)~u\\
\qquad = ((\lambda x:A.t)*t_2\oplus (\lambda x:A.t)*t_3)~u\\
\qquad = (\lambda x:A.(t*t_2)\oplus \lambda x:A.(t*t_3))~u \\
\qquad = (\lambda x:A.(t*t_2)\oplus (t*t_3))~u \\
\qquad= (t[u/x]*t_2)\oplus (t[u/x]*t_3)

\end{array}
\]

\ \ Based on induction on type B, we have $t[u/x]*(t_2\oplus t_3)= (t[u/x]*t_2)\oplus (t[u/x]*t_3)$, Therefore we prove the case.

\smallskip

\item Case:$t_1$ is of type $(T_1,T_2,..,T_{n})$, $t_2$ and $t_3$ are of base type. Suppose $t_1$ is $(t_1',t_2',..,t_{n}')$.

\ \ Then
\[
\begin{array}{llll}
t_1*(t_2\oplus t_3)\\
\qquad=(t_1',t_2',..,t_{n}')*(t_2\oplus t_3)\\
\qquad=(t_1'*(t_2\oplus t_3),t_2'*(t_2\oplus t_3),..,t_{n}'*(t_2\oplus t_3))
\end{array}
\]

\[
\begin{array}{llll}

(t_1*t_2)\oplus (t_1*t_3)\\
\qquad=(t_1',t_2',..,t_{n}')*t_2\oplus (t_1',t_2',..,t_{n}')*t_3\\
\qquad=(t_1'*t_2,t_2'*t_2,..,t_{n}'*t_2)\oplus (t_1'*t_3,t_2'*t_3,..,t_{n}'*t_3)\\
\qquad=(t_1'*t_2\oplus t_1'* t_3,t_2'*t_2\oplus t_2'*t_3,..,t_{n}'*t_2\oplus t_{n}'*t_3)
\end{array}
\]

\ \ And based on induction we have  $t_j'*(t_2\oplus t_3) = (t_j'*t_2)\oplus (t_j'*t_3)$, so we have $t_1*(t_2\oplus t_3) = (t_1*t_2)\oplus (t_1*t_3)$.

\smallskip

\item Case:$t_1$ is of type $T_1+T_2$, $t_2$ and $t_3$ are of base type: this case is not possible because the righthand term is not well-typed.

\item Case:$t_1$ is of type $(T_1,T_2,..,T_{n})$, $t_2$ and $t_3$ are of type $(T_1',T_2',..,T_{n}')$. Suppose $t_1:(t_{11}',t_{12}',..,t_{1n}')$,$t_2:(t_{21}',t_{22}',..,t_{2n}')$ and $t_3=(t_{31}',t_{32}',..,t_{3n}')$.

\ \ Then

\[
\begin{array}{llll}
t_1*(t_2\oplus t_3)\\
\qquad=(t_{11}',t_{12}',..,t_{1n}')*((t_{21}',t_{22}',..,t_{2n}')\oplus (t_{31}',t_{32}',..,t_{3n}'))\\
\qquad=t_{11}'*(t_{21}'\oplus t_{31}')\oplus t_{12}'*(t_{22}'\oplus t_{32}')\oplus ... \oplus t_{1n}'*(t_{2n}'\oplus t_{3n}')
\end{array}
\]

\ \ And we have

\[
\begin{array}{llll}
(t_1*t_2)\oplus (t_1*t_3)\\
\qquad=(t_{11}',t_{12}',..,t_{1n}')*(t_{21}',t_{22}',..,t_{2n}')\oplus (t_{11}',t_{12}',..,t_{1n}')*(t_{31}',t_{32}',..,t_{3n}')\\
\qquad=((t_{11}'*t_{21}')\oplus (t_{11}'*t_{31}'))\oplus((t_{12}'*t_{22}')\oplus (t_{12}'*t_{32}'))\oplus ... \oplus ((t_{1n}'*t_{2n}')\oplus (t_{1n}'*t_{3n}'))
\end{array}
\]

\ \ Based on induction we have $\forall j,t_{1j}'*(t_{2j}'\oplus t_{3j}')=((t_{1j}'*t_{2j}')\oplus (t_{1j}'*t_{3j}'))$, and using Lemma \ref{EqAdd} $t_1 =t_1',t_2=t_2'$, then $t_1\oplus t_2 = t_1'\oplus t_2' $ we prove the case.

\end{itemize}
\end{proof}

\begin{lemma}
\rm

If $(t_1\ominus t_2)\oplus(t_2 \ominus t_3)$ and $t_1\ominus t_3$ are weak-normalizable, then $(t_1\ominus t_2)\oplus(t_2 \ominus t_3) = t_1\ominus t_3$
\end{lemma}

\begin{proof}

If $t_1$, $t_2$ or $t_3$ is not closed, then we just substitute them to be closed. Because of the confluence and normalization property, we can assume that $t_1$, $t_2$ and $t_3$ are all normal forms.

Then we make induction on \xmodify{types}{type} of $t_1$.

\begin{itemize}
\item Case base type

Then because on base type, we require that $(t_1\ominus t_2)\oplus(t_2 \ominus t_3) = t_1\ominus t_3$, Thus we \xmodify{prove}{have proven} the case.

\smallskip

\item Case A$\rightarrow$B

\ \ Then we need to prove that $\forall u,((t_1\ominus t_2)\oplus(t_2 \ominus t_3)) u = (t_1\ominus t_3) u$.

\ \ Then we can suppose that $t_1$, $t_2$ and $t_3$ are of the form $\lambda a:A.t$, if they are not, then we use $\lambda a:A.t_i~a$ to substitute for $t_i$.

\ \ Then we have

\[
\begin{array}{llll}
((t_1\ominus t_2)\oplus(t_2 \ominus t_3)) u\\
\qquad= ((\lambda a:A.t_1'\ominus \lambda a:A.t_2')\oplus(\lambda a:A.t_2' \ominus \lambda a:A.t_3')) u\\
\qquad= \lambda a:A.((t_1'\ominus t_2')\oplus(t_2' \ominus t_3')) u\\
\qquad= ((t_1'[u/a]\ominus t_2'[u/a])\oplus(t_2'[u/a] \ominus t_3'[u/a]))\\
\qquad= ((\lambda a:A.t_1'~u\ominus \lambda a:A.t_2'~u)\oplus(\lambda a:A.t_2'~u \ominus \lambda a:A.t_3'~u))\\
\qquad = ((t_1~u\ominus t_2~u)\oplus(t_2~u \ominus t_3~u))
\end{array}
\]

\ \ And similarly we have $(t_1\ominus t_3) u$ = $(t_1~u\ominus t_3~u)$\xmodify{none}{.}

\ \ \xmodify{Base}{Based} on induction on type $\basetype$, we have $(t_1~u\ominus t_3~u) = ((t_1~u\ominus t_2~u)\oplus(t_2~u \ominus t_3~u))$\xmodify{none}{.}

\ \ Thus we \xmodify{prove}{have proven} the case.

\smallskip

\item Case $(T_1,T_2,...,T_n)$

\ \ Let's suppose $t_1$ to be $(t_{11},t_{12},...,t_{1n})$, $t_2$ to be $(t_{21},t_{22},...,t_{2n})$ and $t_3$ to be $(t_{31},t_{32},...,t_{3n})$.

\ \ Then we have
\[
\begin{array}{llll}
((t_1\ominus t_2)\oplus(t_2 \ominus t_3)) \\
\qquad=(((t_{11}\ominus t_{21})\oplus(t_{21} \ominus t_{31})),...,((t_{1n}\ominus t_{2n})\oplus(t_{2n} \ominus t_{3n})))
\end{array}
\]

\ \ And
\[
\begin{array}{llll}
(t_1\ominus t_3)\\
\qquad=((t_{11} \ominus t_{31}),...,(t_{1n} \ominus t_{3n}))
\end{array}
\]
\ \ Base on induction on type $T_i$, we have $(t_{1i}\ominus t_{2i})\oplus(t_{2i} \ominus t_{3i}) = (t_{1i} \ominus t_{3i})$
\ \ Thus we prove the case.
\xmodify{item}{}
\item Case $T_1+T_2$: This case is not possible because it is not well-typed.

\end{itemize}

\ \ Thus we \xmodify{prove}{have proven} the theorem.

\end{proof}

\end{document}